%% file: serial_speakers.tex
\documentclass[10pt, a4paper]{article}
\usepackage{lrec}
\usepackage{graphicx}
\usepackage{tabularx}
\usepackage{soul}

\usepackage{epstopdf}
\usepackage[utf8]{inputenc}
\usepackage{color}

\usepackage[bookmarks=true, bookmarksnumbered=true, bookmarksopen=true, unicode=true, colorlinks=true, 
linkcolor=blue,citecolor=blue,filecolor=blue,urlcolor=blue, pdfstartview=FitH]
{hyperref}

\usepackage{xstring}

\usepackage{subfig}

\usepackage[table]{xcolor}
\usepackage{booktabs}
\usepackage[super]{nth}
\usepackage{amssymb}
\usepackage{pifont}
\newcommand{\cmark}{\ding{51}}
\newcommand{\xmark}{\ding{55}}

\title{Serial Speakers: a Dataset of TV Series}

\name{Xavier Bost, Vincent Labatut, Georges Linar\`es}

\address{Orkis, Laboratoire Informatique d'Avignon \\
         13290 Aix-en-Provence, France, 84000 Avignon, France \\
         \{firstname.lastname\}@univ-avignon.fr\\}

       \abstract{For over a decade, \textsc{tv} series have been
         drawing increasing interest, both from the audience and from
         various academic fields. But while most viewers are hooked on
         the continuous plots of \textsc{tv} serials, the few
         annotated datasets available to researchers focus on
         standalone episodes of classical \textsc{tv} series. We aim
         at filling this gap by providing the multimedia/speech
         processing communities with \textit{Serial Speakers}, an
         annotated dataset of 161 episodes from three popular American
         \textsc{tv} serials: \textit{Breaking Bad}, \textit{Game of
           Thrones} and \textit{House of Cards}. \textit{Serial
           Speakers} is suitable both for investigating multimedia
         retrieval in realistic use case scenarios, and for addressing
         lower level speech related tasks in especially challenging
         conditions. We publicly release annotations for every speech
         turn (boundaries, speaker) and scene boundary, along with
         annotations for shot boundaries, recurring shots, and
         interacting speakers in a subset of episodes. Because of
         copyright restrictions, the textual content of the speech
         turns is encrypted in the public version of the dataset, but
         we provide the users with a simple online tool to recover the
         plain text from their own
         subtitle files.\\
         \newline \\
         \newline \textcolor{red}{\textbf{Cite as:}\\Xavier~Bost, Vincent~Labatut, Georges~Linar\`es.\\\textit{Serial Speakers: a Dataset of TV Series.}\\12th International Conference on Language Resources and Evaluation (LREC 2020).}\\
         \newline
         \textbf{Note:} This is a slightly extended version of the
         official LREC paper, including additional statistics for the
         final, eighth season of \textit{Game of Thrones}, annotated
         after the paper was submitted.
       }

\begin{document}

\maketitleabstract

\section{Introduction}
\label{sec:intro}

\begin{figure}[!h]
  \subfloat[Viewing frequency (\%).
  \label{subfig:viewing_frequency}]{%
    \includegraphics[width=0.53\linewidth]{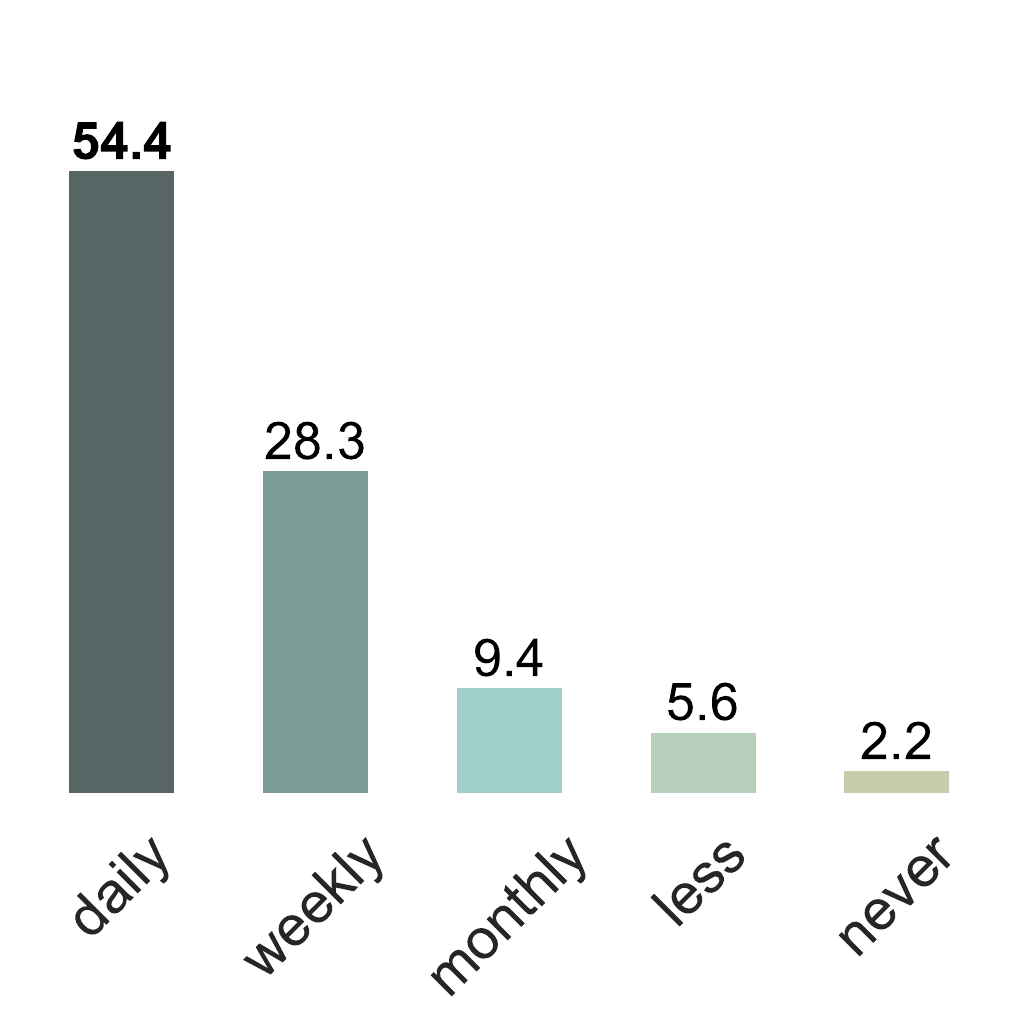}
  }
  \hfill
  \subfloat[Viewing media (\%).
  \label{subfig:viewing_media}]
  {%
    \includegraphics[width=0.43\linewidth]{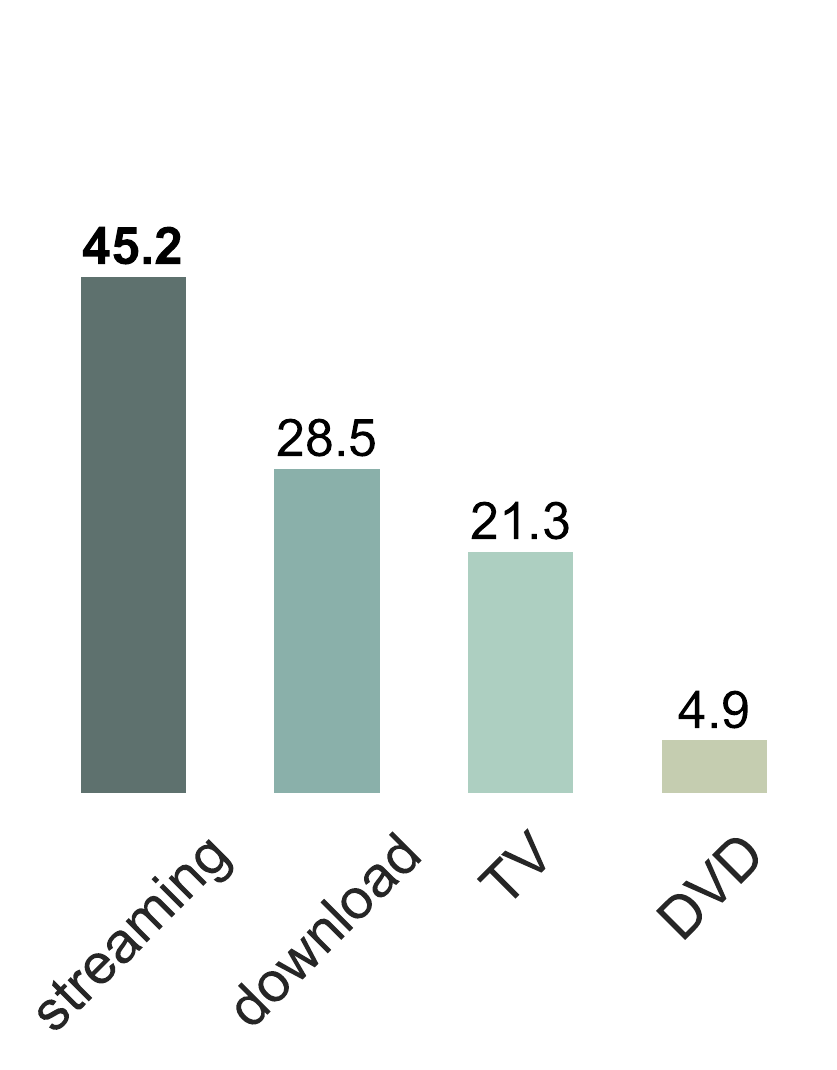}
  }
  \caption{TV series, viewing conditions.}
  \label{fig:viewing_conditions}
\end{figure}

For over a decade now, \textsc{tv} series have been drawing increasing
attention. In 2019, the final season of \textit{Game of Thrones}, one
of the most popular \textsc{tv} shows these past few years, has
averaged 44.2 million viewers per episode; many \textsc{tv} series
have huge communities of fans, resulting in numerous online
crowdsourced resources, such as
wikis\footnote{\href{http://gameofthrones.wikia.com/wiki/Game_of_Thrones_Wiki}{gameofthrones.wikia.com/wiki/Game\_of\_Thrones\_Wiki}},
dedicated
forums\footnote{\href{https://asoiaf.westeros.org/index.php?/forum/31-game-of-thrones-the-hbo-tv-series}{asoiaf.westeros.org/index.php?/forum}},
and \textit{YouTube} channels. Long dismissed as a minor genre by the
critics, some recent \textsc{tv} series also received critical acclaim
as a unique space of creativity, able to attract even renowned
full-length movie directors, such as Jane Campion, David Fincher or
Martin Scorsese. Nowadays, \textsc{tv} series have their own
festivals\footnote{In France, \href{https://seriesmania.com/en}{Series
    Mania}.}. For more than half of the people\footnote{194
  individuals, mostly students from our university, aged 23.12 $\pm$
  5.73.} we polled in the survey reproduced in \cite{Bost2016},
watching \textsc{tv} series is a daily occupation, as can be seen on
Fig.~\ref{subfig:viewing_frequency}.

Such a success is probably related to the cultural changes caused by
modern media: high-speed internet connections led to unprecedented
viewing opportunities. As shown on Fig.~\ref{subfig:viewing_media},
television is no longer the main channel used to watch ``\textsc{tv}''
series: most of the time, streaming and downloading services are
preferred to television.

\begin{figure}[!h]
  \subfloat[Season viewing time.
  \label{subfig:season_viewing_time}]{%
    \includegraphics[width=0.43\linewidth]{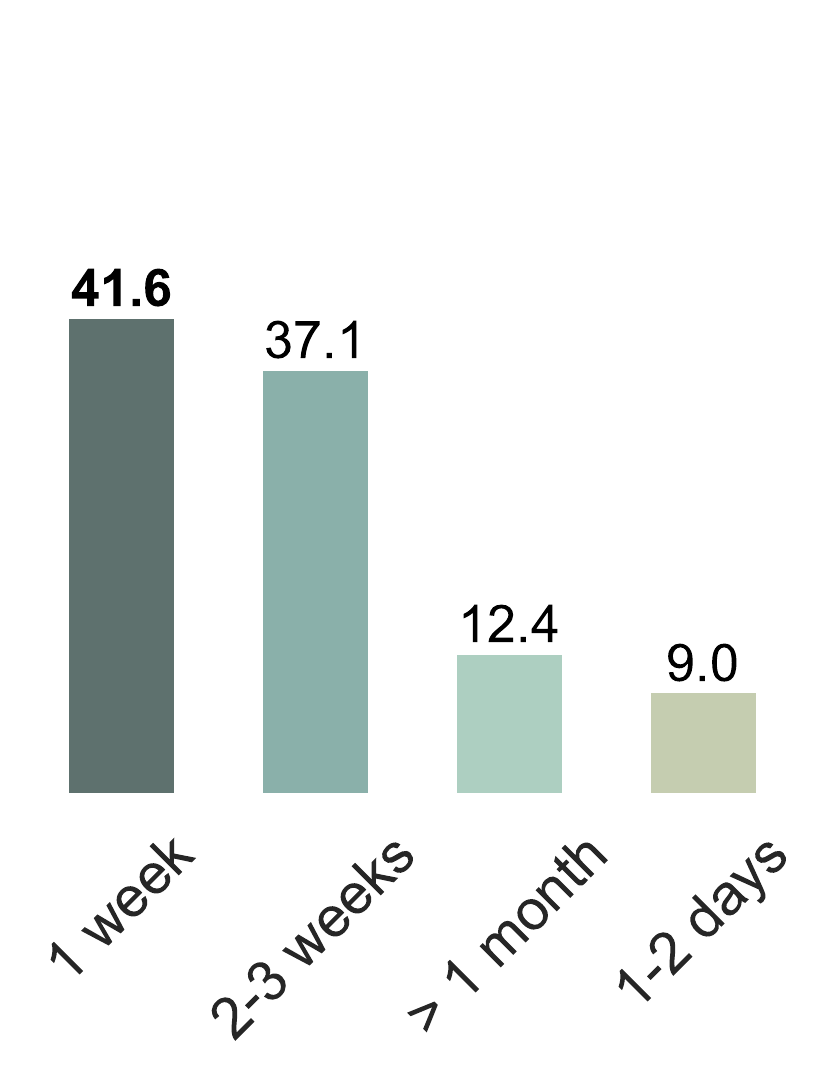}
  }
  \hfill
  \subfloat[Favorite genre.
  \label{subfig:favorite_genre}]
  {%
    \includegraphics[width=0.43\linewidth]{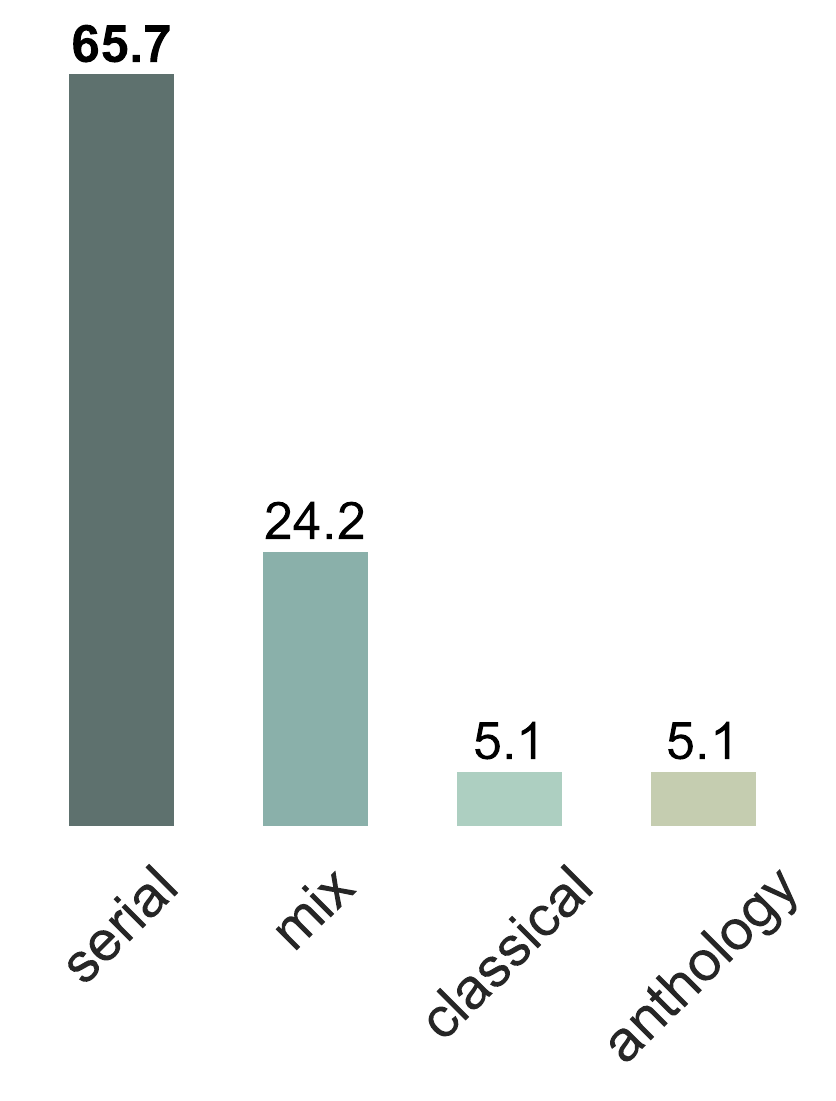}
  }
  \caption{TV series, season viewing time; favorite genre.}
  \label{fig:viewing_time_favorite_genre}
\end{figure}

Unlike television, streaming and downloading platforms give control to
the user, not only over the contents he may want to watch, but also
over the viewing frequency. As a consequence, the typical dozen of
episodes a \textsc{tv} series season contains is often watched over a
much shorter period of time than the usual two months it is being
broadcast on television. As can be seen on
Fig.~\ref{subfig:season_viewing_time}, for almost 80\% of the
people we polled, watching a \textsc{tv} series season (about 10 hours
in average) never takes more than a few weeks. As a major consequence,
\textsc{tv} series seasons, usually released once a year, are not
watched in a continuous way.

For some types of \textsc{tv} series, discontinuous viewing is
generally not a major issue. Classical \textsc{tv} series consist of
self-contained episodes, only related with one another by a few
recurring protagonists. Similarly, anthologies contain standalone
units, either episodes (\textit{e.g.} \textit{The Twilight Zone}) or
seasons (\textit{e.g.} \textit{True detective}), but without recurring
characters. However, for \textsc{tv} \textit{serials}, discontinuous
viewing is likely to be an issue: \textsc{tv} serials (\textit{e.g.}
\textit{Game of Thrones}) are based on highly continuous plots, each
episode and season being narratively related to the previous ones.

Yet, as reported on Fig.~\ref{subfig:favorite_genre}, \textsc{tv} serials
turn out to be much more popular than classical \textsc{tv} series:
nearly 2/3 of the people we polled prefer \textsc{tv} serials to the
other types, and 1/4 are more inclined to a mix between the classical
and serial genres, each episode developing its own plot but also
contributing to a secondary, continuous story.

As a consequence, viewers are likely to have forgotten to some extent
the plot of \textsc{tv} serials when they are, at last, about to know
what comes next: nearly 60\% of the people we polled feel the need to
remember the main events of the plot before viewing the new season of
a \textsc{tv} serial. Such a situation, quite common, provides
multimedia retrieval with remarkably realistic use cases.

A few works have been starting to explore multimedia retrieval for
\textsc{tv} series. \newcite{Tapaswi2014a} investigate ways of automatically
building
\textsc{xkcd}-style\footnote{\href{https://xkcd.com/657}{xkcd.com/657}}
visualizations of the plot of \textsc{tv} series episodes based on the
interactions between onscreen characters. \newcite{Ercolessi2012a}
explore plot de-interlacing in \textsc{tv} series based on scene
similarities.  \newcite{Bost2019} made use of automatic extractive
summaries for re-engaging viewers with \textit{Game of Thrones}' plot,
a few weeks before the sixth season was released. \newcite{Roy2014} and
\newcite{Tapaswi2014b} make use of crowdsourced plot synopses which, once
aligned with video shots and/or transcripts, can support high-level,
event-oriented search queries on \textsc{tv} series content.

Nonetheless, most of these works focus either on classical \textsc{tv}
series, or on standalone episodes of \textsc{tv} serials. Due to the
lack of annotated data, very few of them address the challenges
related to the narrative continuity of \textsc{tv} serials. We aim at
filling this gap by providing the multimedia/speech processing
research communities with \textit{Serial Speakers}, an annotated
dataset focusing on three American \textsc{tv} serials:
\textit{Breaking Bad} (seasons 1--5~/~5), \textit{Game of Thrones}
(seasons 1--8~/~8), \textit{House of Cards} (seasons
1--2~/~6). Besides multimedia retrieval, the annotations we provide
make our dataset suitable for lower level tasks in challenging
conditions (Subsection~\ref{subsec:speech_turns}). In this paper,
we first describe the few existing related datasets, before detailing
the main features of our own \textit{Serial Speakers} dataset; we finally
describe the tools we make available to the users for reproducing the
copyrighted material of the dataset.

\section{Related Works}
\label{sec:related_works}

These past ten years, a few commercial \textsc{tv} series have been
annotated for various research purposes, and some of these annotations
have been publicly released. We review here most of the \textsc{tv}
shows that were annotated, along with the corresponding types of
annotations, whenever publicly available.

\textbf{\textit{Seinfeld}} (1989--1998) is an American \textsc{tv}
\textit{situational comedy (sitcom)}. \newcite{Friedland2009} rely on
acoustic events to design a navigation tool for browsing episodes
publicly released during the \textsc{acm} Multimedia 2009 Grand
Challenge.

\textbf{\textit{Buffy the Vampire Slayer}} (1997--2001) is an American
\textit{supernatural drama} \textsc{tv} series. This show was mostly
used for character naming \cite{Everingham2006}, face tracking and
identification \cite{Bauml2013}, person identification
\cite{Bauml2014}, \cite{Tapaswi2015a}, story visualization
\cite{Tapaswi2014a}, and plot synopses alignment
\cite{Tapaswi2014b}\footnote{Visual (face tracks and identities) and
  linguistic (video alignment with plot synopses) annotations of the
  fifth season can be found at
  \href{https://cvhci.anthropomatik.kit.edu/~mtapaswi/projects-mma.html}{cvhci.anthropomatik.kit.edu/mtapaswi/projects-mma.html}}.

\textbf{\textit{Ally McBeal}} (1997--2002) is an American
\textit{legal comedy-drama} \textsc{tv} series. The show was annotated
for performing scene segmentation based on speaker diarization
\cite{Ercolessi2011} and speech recognition \cite{Bredin2012}, plot
de-interlacing \cite{Ercolessi2012a}, and story visualization
\cite{Ercolessi2012b}\footnote{Annotations (scene/shot boundaries,
  speaker identity) of the first four episodes are available at
  \href{http://herve.niderb.fr/data/ally\_mcbeal}{herve.niderb.fr/data/ally\_mcbeal}}.

\textbf{\textit{Malcom in the Middle}} (2000--2006) is an American
\textsc{tv} \textit{sitcom}. Seven episodes were annotated for story
de-interlacing \cite{Ercolessi2012a} and visualization
\cite{Ercolessi2012b} purposes.

\textbf{\textit{The Big Bang Theory}} (2007--2019) is also an American
\textsc{tv} \textit{sitcom}. Six episodes were annotated for the same
visual tasks as those performed on \textit{Buffy the Vampire Slayer}:
face tracking and identification \cite{Bauml2013}, person
identification \cite{Bauml2014}, \cite{Tapaswi2015a}, and story
visualization \cite{Tapaswi2014a}. \newcite{Tapaswi2012} also focus on
speaker identification and provide audiovisual annotations for these
six
episodes\footnote{\href{https://cvhci.anthropomatik.kit.edu/~mtapaswi/projects-personid.html}{cvhci.anthropomatik.kit.edu/mtapaswi/projectspersonid.html}}. In
addition to these audiovisual annotations, \newcite{Roy2014} publish in
the \textsc{tvd} dataset other crowdsourced, linguistically oriented
resources, such as manual transcripts, subtitles, episode outlines and
textual
summaries\footnote{\href{http://tvd.niderb.fr/}{tvd.niderb.fr/}}.

\begin{table*}[!h]
  \begin{center}
    \input{speech_features}
    \caption{Speech features.}
    \label{tab:speech_features}
  \end{center}
\end{table*}

\textbf{\textit{Game of Thrones}} (2011--2019) is an American
\textit{fantasy drama}. \newcite{Tapaswi2014a} make use of annotated face
tracks and face identities in the first season (10 episodes). In
addition, \newcite{Tapaswi2015b} provide the ground truth alignment
between the first season of the \textsc{tv} series and the books it is
based
on\footnote{\href{https://cvhci.anthropomatik.kit.edu/~mtapaswi/projects-book\_align.html}{cvhci.anthropomatik.kit.edu/mtapaswi/projectsbook\_align.html}}. For
a subset of episodes, the \textsc{tvd} dataset provides crowdsourced
manual transcripts, subtitles, episode outlines and textual summaries.

As can be seen, many of these annotations target vision-related
tasks. Furthermore, little attention has been paid to \textsc{tv}
serials and their continuous plots, usually spanning several
seasons. Instead, standalone episodes of sitcoms are overrepresented. And
finally, even when annotators focus on \textsc{tv} serials
(\textit{Game of Thrones}), the annotations are never provided for
more than a single season. Similar to the computer vision
\textsc{accio} dataset for the series of \textit{Harry Potter}
movies \cite{Ghaleb2015}, our \textit{Serial Speakers} dataset aims in contrast at
providing annotations of several seasons of
\textsc{tv} serials, in order to address both the realistic multimedia
retrieval use cases we detailed in Section~\ref{sec:intro}, and lower
level speech processing tasks in unusual, challenging conditions.

\begin{table}[!h]
  \begin{center}
    \input{gen_features}
    \caption{Duration of the video recordings.}
    \label{tab:gen_features}
  \end{center}
\end{table}

\section{Description of the Dataset}
\label{sec:dataset_description}

Our \textit{Serial Speakers} dataset consists of 161 episodes from three
popular \textsc{tv} serials:

\paragraph{\textit{Breaking Bad}} (denoted hereafter \textsc{bb}),
released between 2008 and 2013, is categorized on \textit{Wikipedia} as a \textit{crime
  drama}, \textit{contemporary western} and a \textit{black comedy}.
We annotated 62 episodes (seasons 1--5) out of 62.

\paragraph{\textit{Game of Thrones}} (\textsc{got}) has been introduced above in
Section~\ref{sec:related_works} We annotated 73 episodes (seasons
1--8) out of 73.

\paragraph{\textit{House of Cards}} (\textsc{hoc}) is a
\textit{political drama}, released between 2013 and 2018. We annotated
26 episodes (seasons 1--2) out of 73.

Overall, the total duration of the video recordings amounts to $\simeq$
135 hours (135:34:27). Table~\ref{tab:gen_features} details for every
season of each of the three \textsc{tv} serials the duration of the
video recordings, expressed in ``HH:MM:SS'', along with the
corresponding number of episodes (in parentheses).

\subsection{Speech Turns}
\label{subsec:speech_turns}

As in any full-length movie, speech is ubiquitous in \textsc{tv}
serials. As reported in Table~\ref{tab:speech_features}, speech
coverage in our dataset ranges from 35\% to 46\% of the video
duration, depending on the \textsc{tv} series, for a total amount of
about 51 hours. As can be seen, speech coverage is much more important
(46\%) in~\textsc{hoc} than in \textsc{bb} and \textsc{got}
(respectively 38\% and 35\%). As a political drama, \textsc{hoc} is
definitely speech oriented, while the other two series also contain
action scenes. Interestingly, speech coverage in~\textsc{got} tends to
decrease over the 8 seasons, especially from the fifth one. The first
seasons turn out to be relatively faithful to the book series they are
based on, while the last ones tend to depart from the original
novel. Moreover, with increasing financial means, \textsc{got}
progressively moved to a pure fantasy drama, with more action scenes.

The basic speech units we consider in our dataset are \textit{speech
  turns}, graphically signaled as sentences by ending punctuation
signs. Unlike speaker turns, two consecutive speech turns may
originate in the same speaker.

\begin{figure*}[!h]
  \subfloat[Speech turns, duration distribution.
  \label{subfig:speech_turns_dur}]{%
    \includegraphics[width=0.36\linewidth]{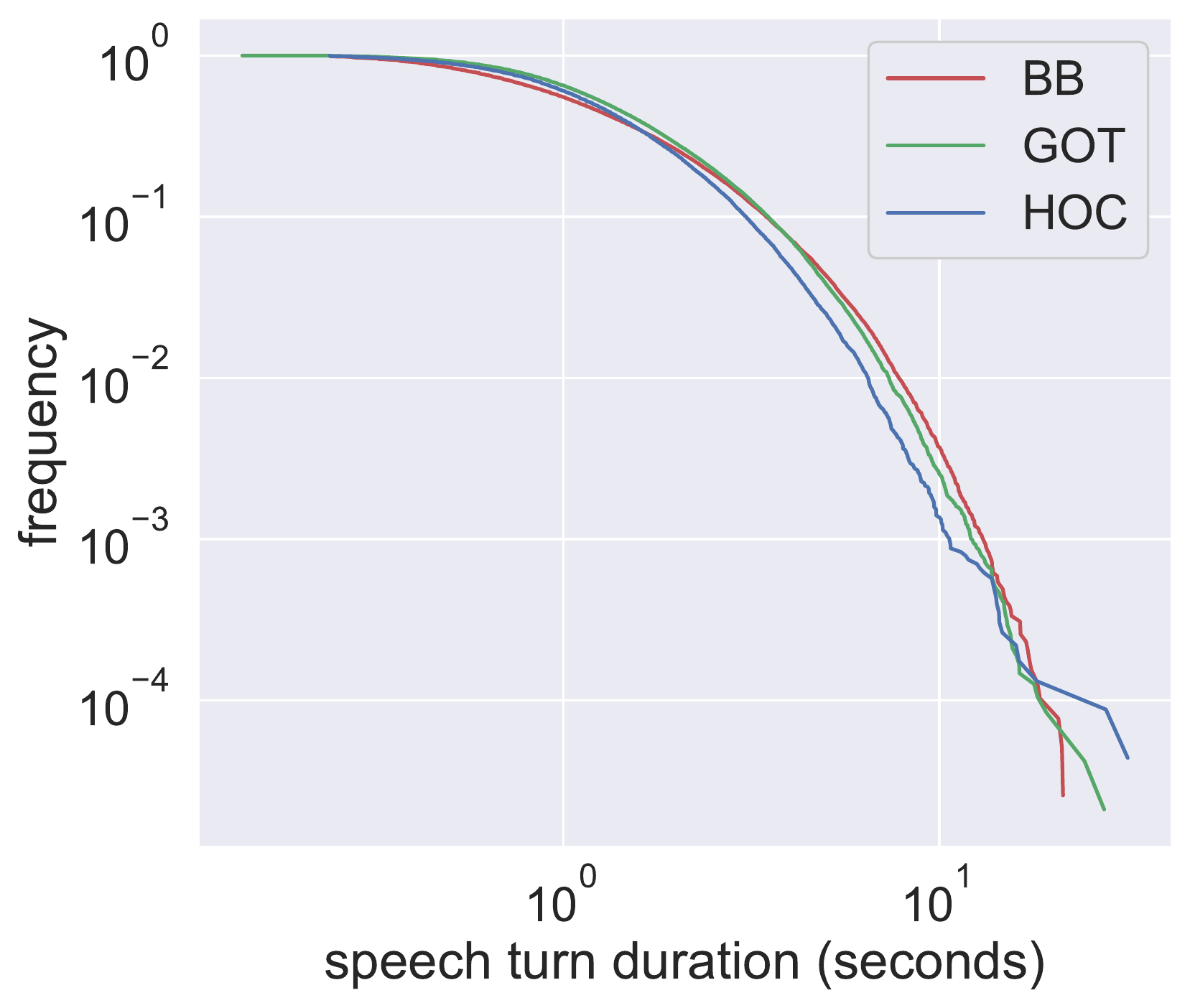}
  }
  \hfill
  \subfloat[Speaking time distribution.
  \label{subfig:speaker_speaking_time}]
  {%
    \includegraphics[width=0.36\linewidth]{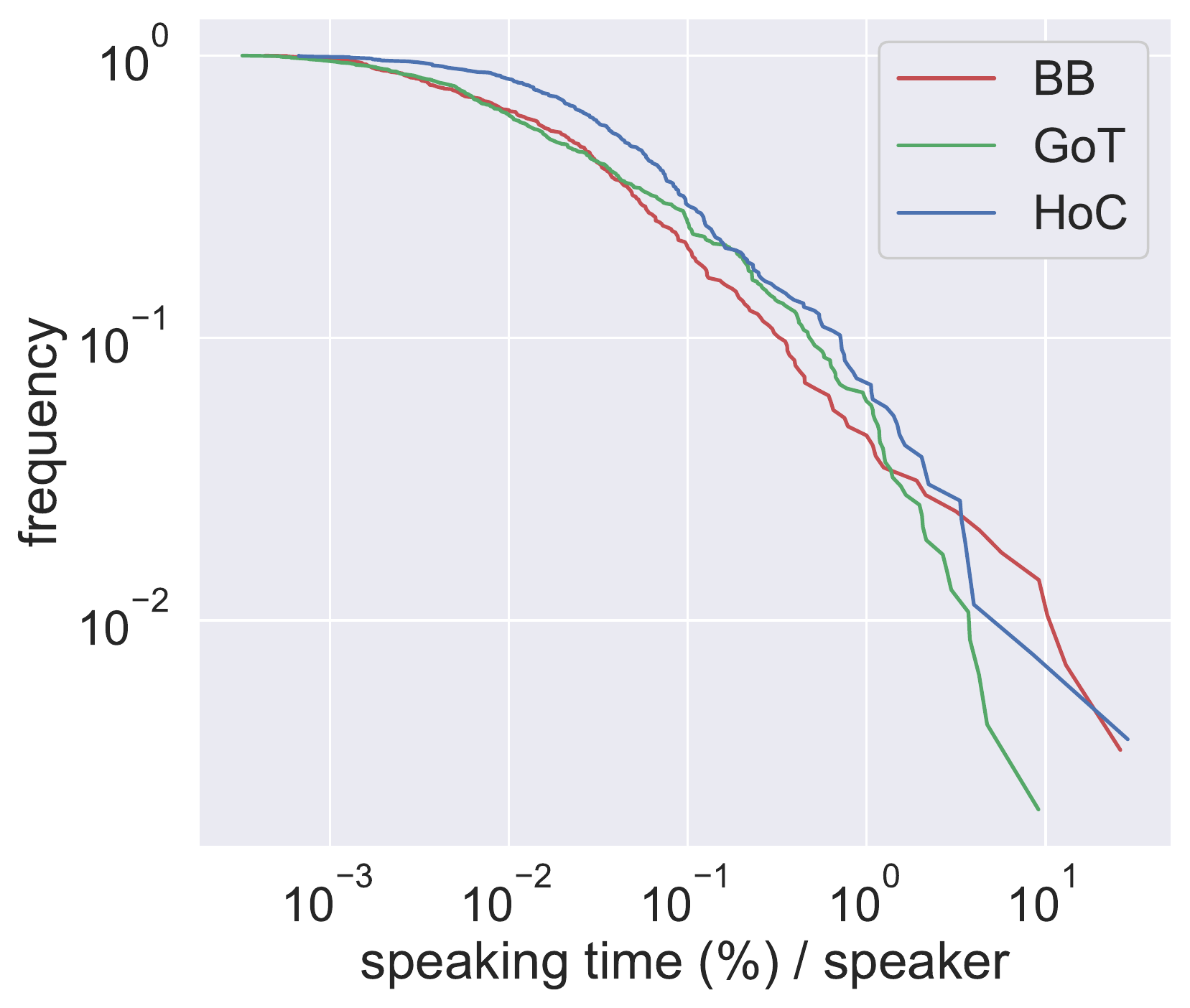}
  }
  \caption{Speech turns duration and speaking time/speaker.}
  \label{fig:viewing_time_favorite_genre}
\end{figure*}

\paragraph{Boundaries.}

The boundaries (starting and ending points) of every speech turn are
annotated. During the annotation process, speech turns were first
based on raw subtitles, as retrieved by applying a standard
\textsc{ocr} tool to the commercial \textsc{dvd}s. Nonetheless,
subtitles do not always correspond to speech turns in a one-to-one way:
long speech turns usually span several consecutive subtitles;
conversely, a single subtitle may contain several speech turns,
especially in case of fast speaker change. We then applied simple
merging/splitting rules to recover the full speech turns from the
subtitles, before refining their boundaries by using the forced
alignment tool described in \cite{Mcauliffe2017}. The resulting
boundaries were systematically inspected and manually adjusted
whenever necessary. Such annotations make our dataset suitable for the
\textit{speech/voice activity detection} task.

Overall, as reported in Table~\ref{tab:speech_features}, the dataset
contains 109,366 speech turns. Speech turns are relatively short: the
median speech turn duration amounts to 1.3 seconds for \textsc{got},
1.2 for \textsc{hoc}, and only 1.1 for \textsc{bb}.

As can be seen on Fig.~\ref{subfig:speech_turns_dur}, the statistical
distribution of the speech turns duration, here plotted on a log-log
scale as a complementary cumulative distribution function, seems to
exhibit a heavy tail in all three cases. This is confirmed more
objectively by applying the statistical testing procedure proposed by
\newcite{Clauset2009}, which shows these distributions follow power
laws. This indicates that the distribution is dominated by very short
segments, but that there is a non-negligible proportion of very long
segments, too. It also reveals that the mean is not an appropriate
statistic to describe this distribution.

\paragraph{Speakers.}

By definition, every speech turn is uttered by a single speaker. We
manually annotated every speech turn with the name of the
corresponding speaking character, as credited in the cast list of each
\textsc{tv} series episode. A small fraction of the speech segments
(\textsc{bb}: 1.6\%, \textsc{got}: 3\%, \textsc{hoc}: 2.2\%) were
left as unidentified (``unknown'' speaker). In the rare cases of two
partially overlapping speech turns, we decided to cut off the first one at
the exact starting point of the second one to preserve as much as possible its purity.

Overall, as can be seen in Table~\ref{tab:speech_features}, 288
speakers were identified in \textsc{bb}, 468 in \textsc{got} and 264
in \textsc{hoc}. With an average speaking time of 132 seconds by
speaker, \textsc{hoc} contains more speakers than \textsc{got}
(175 seconds/speaker), which in turn contains more speakers than
\textsc{bb} (218 seconds/speaker).

Fig.~\ref{subfig:speaker_speaking_time} shows the distribution of the
speaking time (expressed in percentage of the total speech time) for
all speakers, again plotted on a log-log scale as a complementary
cumulative distribution function. Once again, the speaking time of
each speaker seems to follow a heavy-tailed distribution, with a few
ubiquitous speakers and lots of barely speaking characters. This is
confirmed through the same procedure as before, which identifies three
power laws. If we consider that speaking time captures the strength of
social interactions (soliloquies aside), this is consistent with
results previously published for other types of weighted social
networks~\cite{Li2003a,Barthelemy2005}.

Nonetheless, as can be seen on the figure, the main speakers of
\textsc{got} are not as ubiquitous as the major ones in the other two
series: while the five main protagonists of \textsc{bb} and
\textsc{hoc} respectively accumulate 64.3 and 48.6\% of the total
speech time, the five main characters of \textsc{got} ``only''
accumulate 25.6\%. Indeed, \textsc{got}'s plot, based on a choral
novel, is split into multiple storylines, each centered on one major
protagonist.

\begin{figure}[!h]
  \subfloat[\textsc{bb}
  \label{subfig:bb_spk_corr}]{%
    \includegraphics[width=0.38\linewidth]{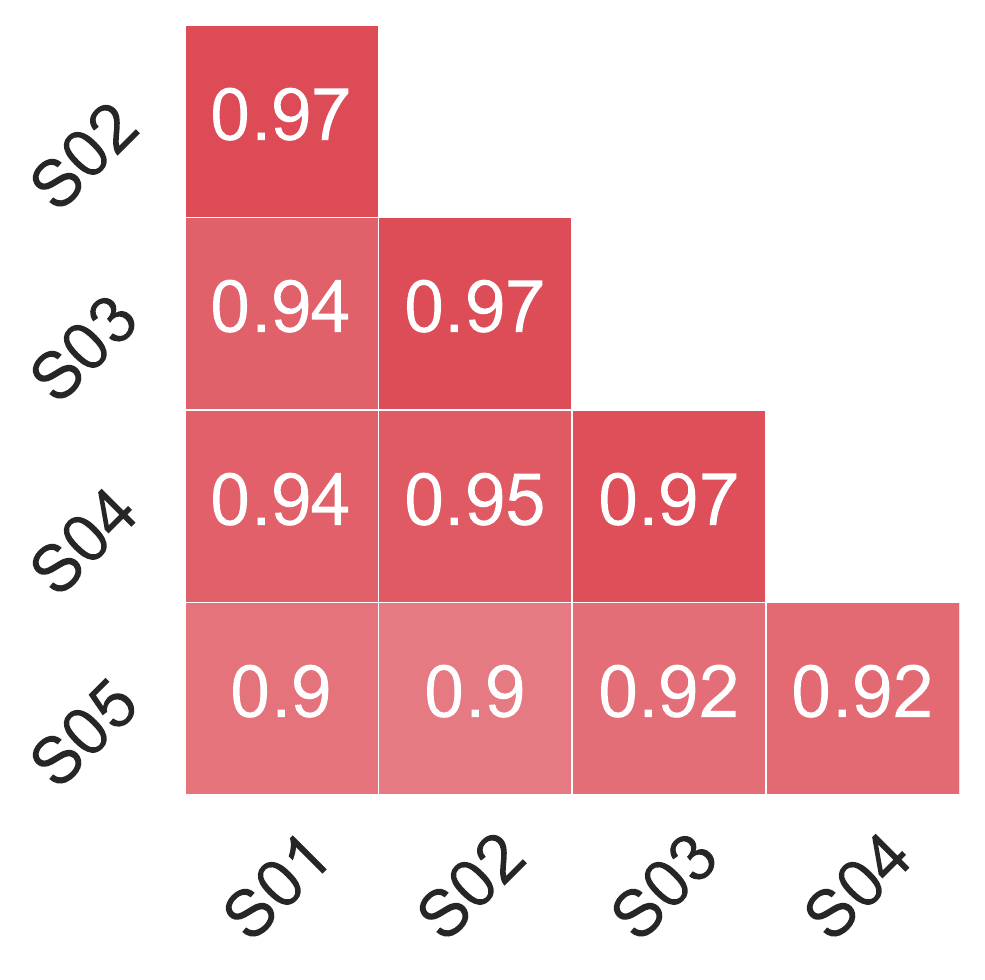}
  }
  \hfill
  \subfloat[\textsc{got}
  \label{subfig:got_spk_corr}]
  {%
    \includegraphics[width=0.6\linewidth]{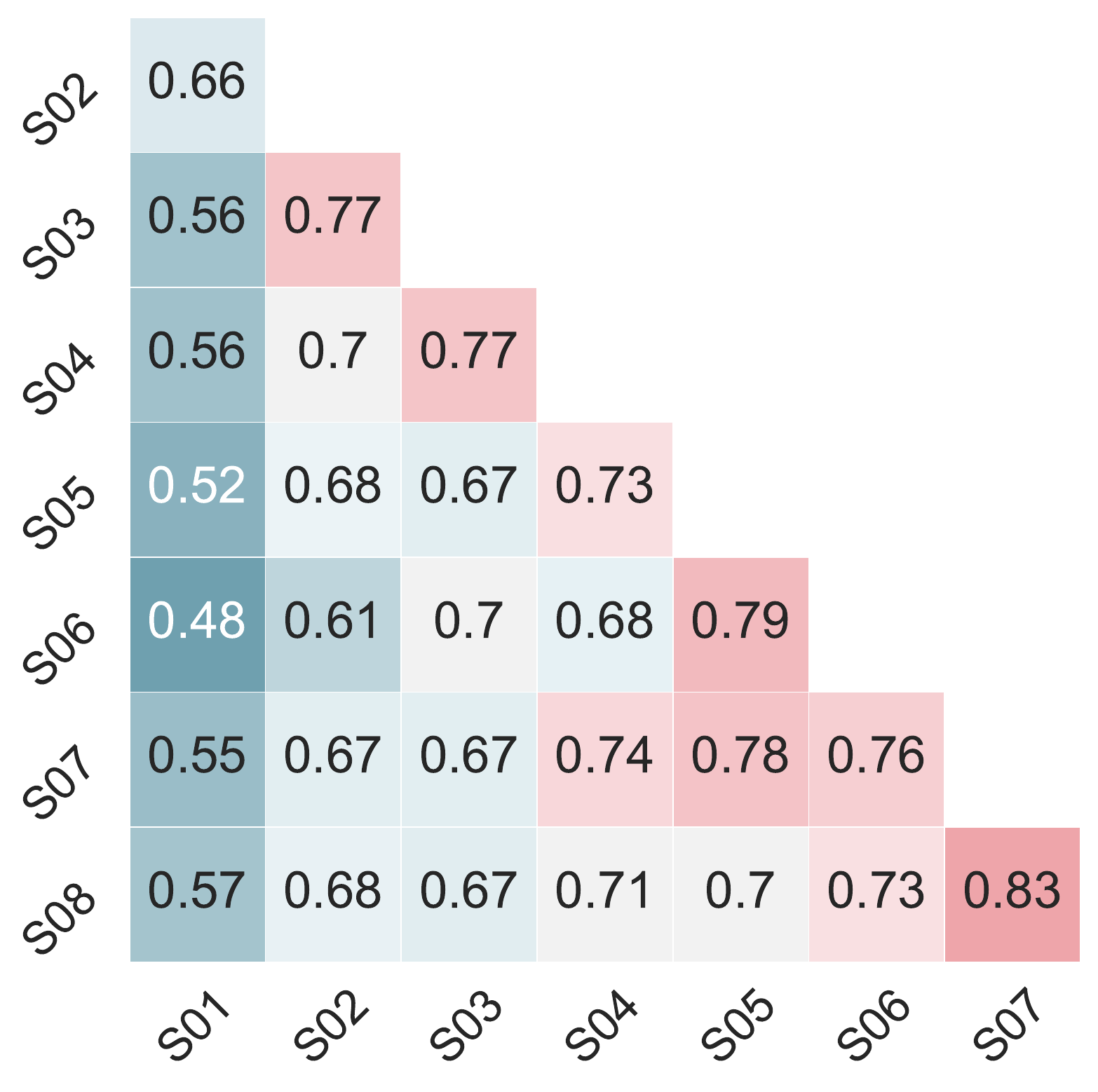}
  }
  \caption{Speakers correlation across seasons.}
  \label{fig:spk_corr}
\end{figure}

Moreover, even major, recurring characters of \textsc{tv} serials are
not always uniformly represented over time. Fig.~\ref{fig:spk_corr}
depicts the lower part of correlation matrices computed between the
speakers involved in every season of \textsc{bb}
(Fig.~\ref{subfig:bb_spk_corr}) and \textsc{got}
(Fig.~\ref{subfig:got_spk_corr}): the distribution of the relative
speaking time of every speaker in each season is first computed,
before the Pearson correlation coefficient is calculated between every
pair of season distribution.

\begin{figure*}[h!]
  \subfloat[\textsc{bb}.
  \label{subfig:bb_network}]{%
    \includegraphics[width=0.32\linewidth]{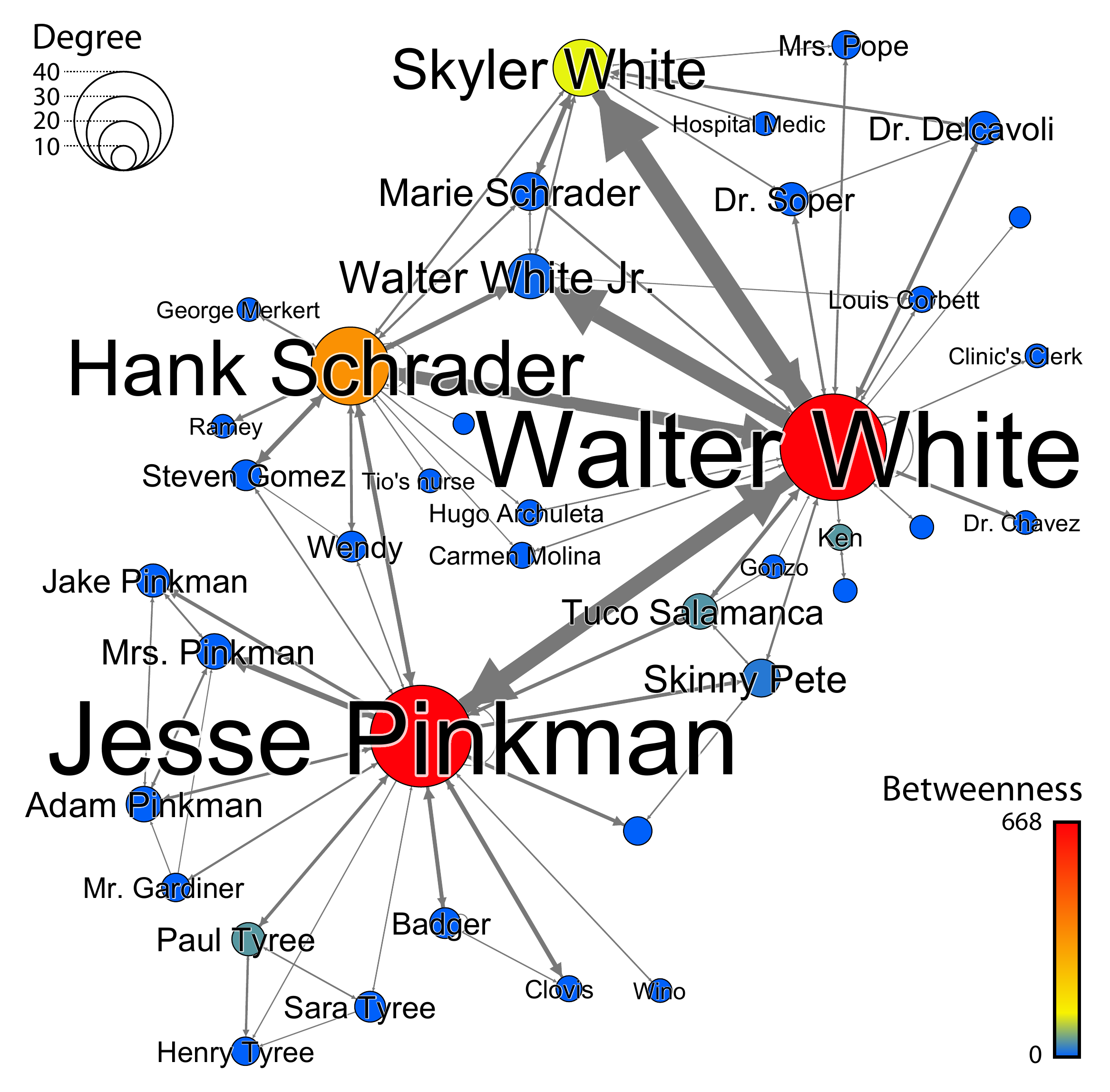}
  }
  \hfill
  \subfloat[\textsc{got}.
  \label{subfig:got_network}]
  {%
    \includegraphics[width=0.32\linewidth]{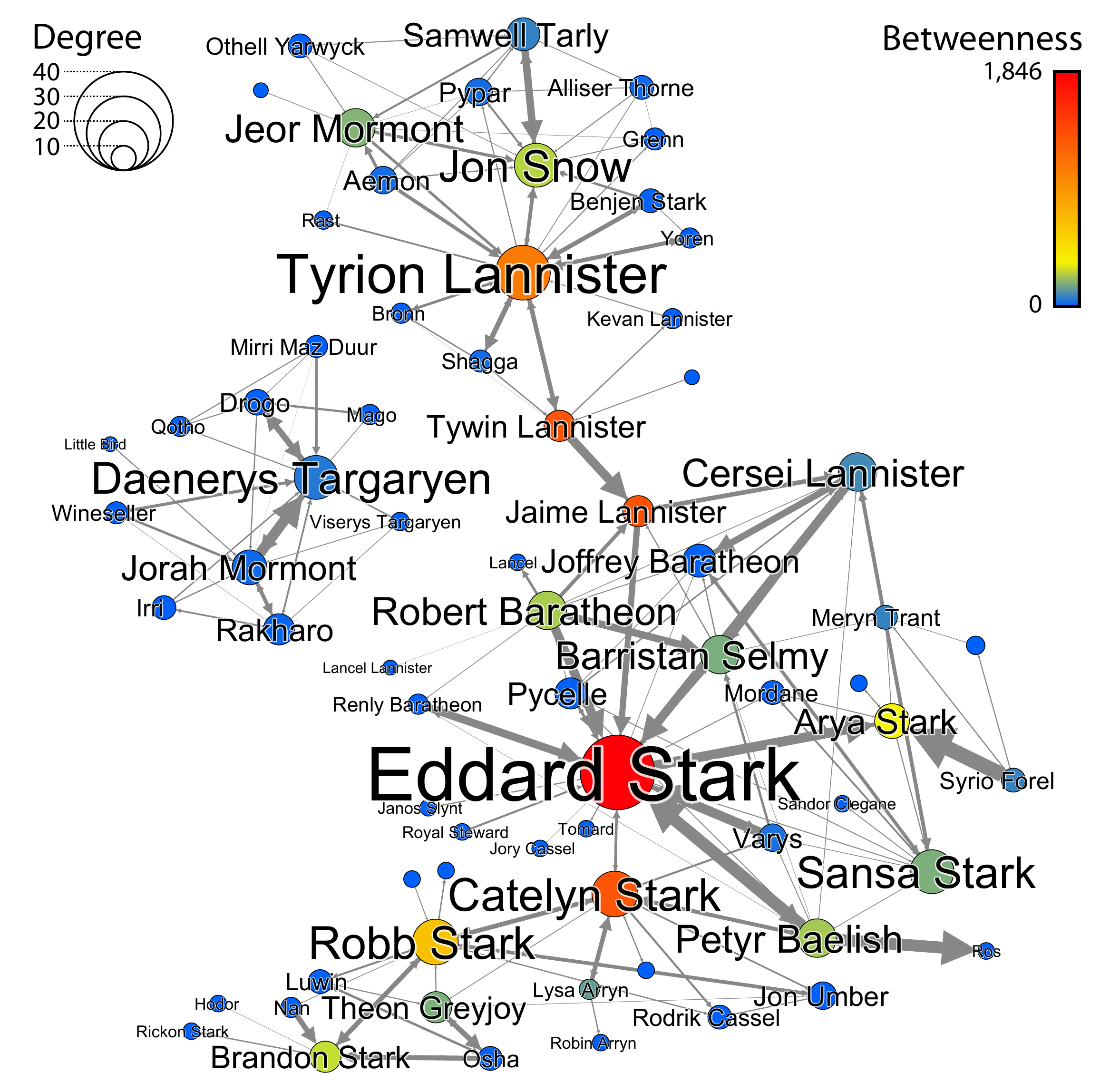}
  }
  \hfill
  \subfloat[\textsc{hoc}.
  \label{subfig:hoc_network}]
  {%
    \includegraphics[width=0.32\linewidth]{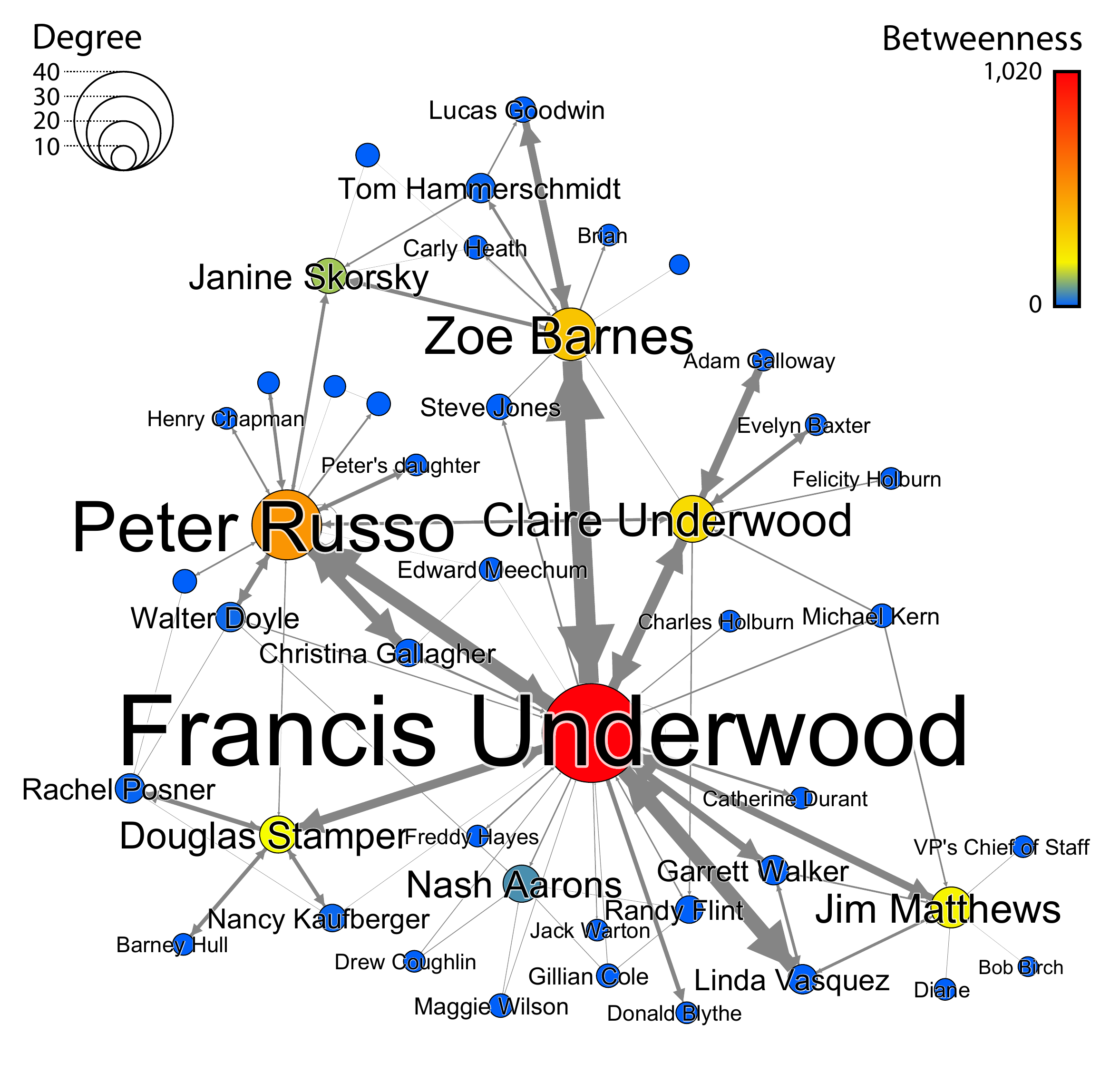}
  }
  \caption{Conversational networks extracted from the annotated episodes. Vertex size and color represent degree and betweenness, respectively.}
  \label{fig:networks}
\end{figure*}

\begin{figure}[!h]
 \begin{center}
   \includegraphics[width=\linewidth]{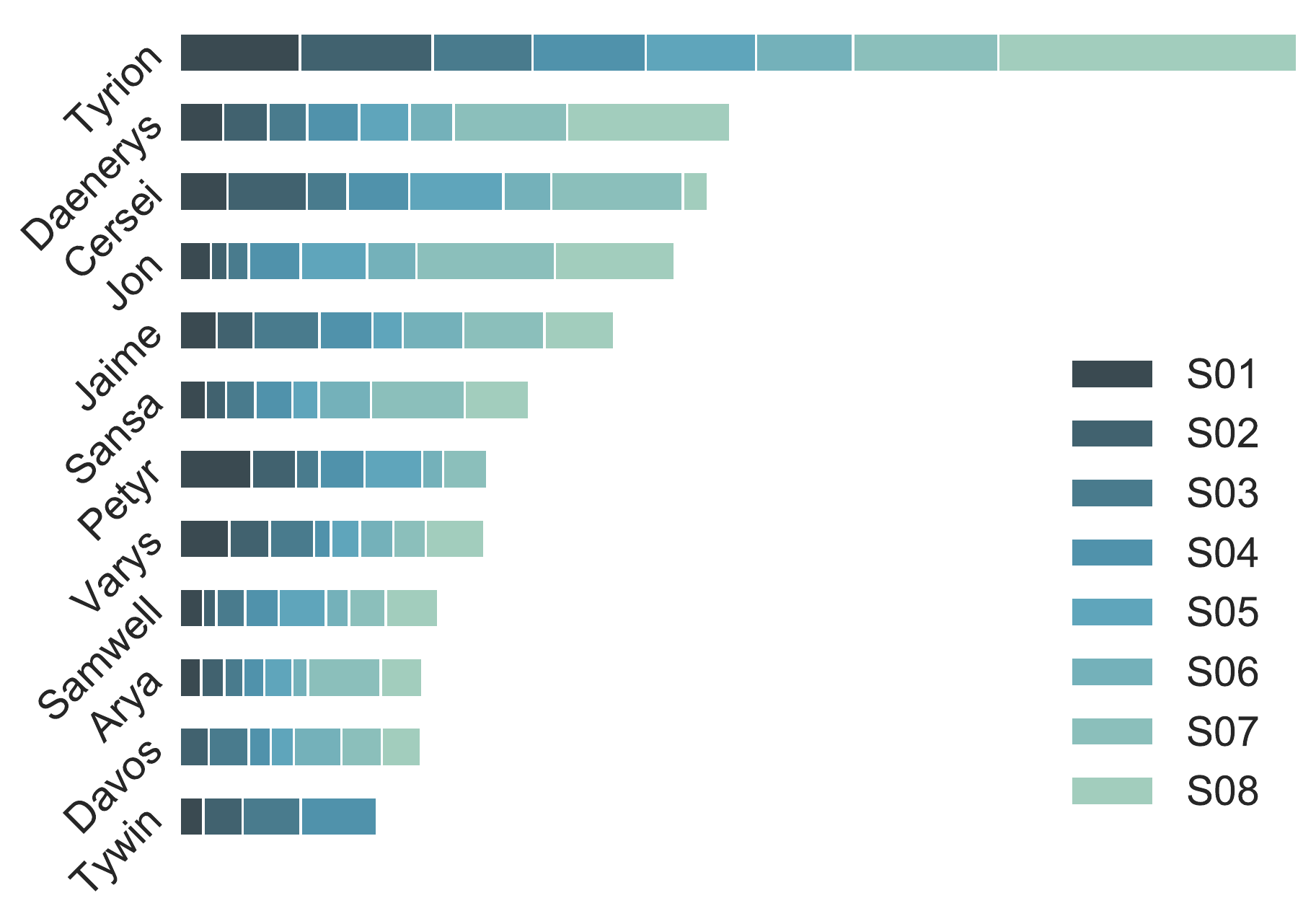} 
   \caption{Relative speaking time over every season of the
     top-12 speakers of \textsc{got}.}
   \label{fig:seasons_speaking_time}
 \end{center}
\end{figure}

As can be seen, the situation is very contrasted, depending on the
\textsc{tv} serial. Whereas the major speakers of \textsc{bb} remain
quite the same over all five seasons (correlation coefficients close
to 1, except for the very last, fifth one, with a few entering new
characters), \textsc{got} exhibits quite lower correlation
coefficients. For instance, the main speakers involved in the first
season turn out to be quite different from the speakers involved in
the other ones (average correlation coefficient with the other seasons
only amounting to 0.56 $\pm$ 0.05). Indeed, \textsc{got} is known for
numerous, shocking deaths of major characters\footnote{See
  \href{https://got.show}{got.show}, for an attempt to automatically
  predict the characters who are the most likely to die
  next.}. Moreover, \textsc{got}'s narrative usually focuses
alternatively on each of its multiple storylines, but may postpone
some of them for an unpredictable time, resulting in uneven speaker
involvement over seasons. Fig.~\ref{fig:seasons_speaking_time} depicts
the relative speaking time in every season of the 12 most active
speakers of \textsc{got}. As can be seen, some characters are barely
present in some seasons, for instance, Jon (rank~\#4) in Season~2, or
even absent, like Tywin (rank~\#12) in Seasons~5--8.

Furthermore, as can be noticed on
Fig.~\ref{fig:seasons_speaking_time}, the relative involvement of most
of these 12 protagonists in Seasons~7--8 is much more important than in
the other ones: indeed, Seasons~7--8 are centered on fewer speakers
(respectively 66 and 50 \textit{vs.} 124.3 $\pm$ 11.8 in average in
the first six ones).

Speaker annotations make our dataset suitable for the speaker
diarization/recognition tasks, but in especially challenging
conditions: first, and as stated in \cite{Bredin2016}, the usual
2-second assumption made for the speech turns by most of the state-of-the-art speaker
diarization systems does no longer stand. Second,
the high number of speakers involved in \textsc{tv} serials,
along with the way their utterances are distributed over time, make
one-step approaches particularly difficult. In such conditions,
multi-stage approaches should be more
effective \cite{Tran2011}. Besides, as noted in \cite{Bredin2016}, the
spontaneous nature of the interactions, the usual background music and
sound effects heavily hurt the performance of standard speaker
diarization/recognition systems \cite{Clement2011}.


\paragraph{Textual content.}

Though not provided with the annotated dataset for obvious copyright
reasons\footnote{Instead, we provide the users with online tools for
  recovering the textual content of the dataset from external subtitle
  files. See Section~\ref{sec:textual_content_generation} for a
  description.}, the textual content of every speech turn has been
revised, based on the output of the \textsc{ocr} tool we used to
retrieve the subtitles. In particular, we restored a few missing
words, mostly for \textsc{bb}, the subtitles sometimes containing some
deletions.

\textsc{bb} contains 229,004 tokens (word occurrences) and 10,152
types (unique words); \textsc{got} 317,840 tokens and 9,275 types; and
\textsc{hoc} 153,846 tokens and 8,508 types.

As the number of tokens vary dramatically from one \textsc{tv} serial
to the other, we used the length-independent \textsc{mtld} measure
\cite{McCarthy2010} to assess the lexical diversity of the three
\textsc{tv} serials. With a value of 88.2 (threshold set to 0.72), the
vocabulary in \textsc{hoc} turns out to be richer than in \textsc{got}
(69.6) and \textsc{bb} (64.5). More speech oriented, \textsc{hoc} also
turns out to exhibit more lexical diversity than the other two series.

\subsection{Interacting Speakers}
\label{subsec:interacting_speakers}

In a subset of episodes, the addressees of every speech turn have been
annotated. Trivial within two-speaker sequences, such a task, even for
annotators, turns out to be especially challenging in more complex
conditions: most of the time, the addressees have to be inferred both
from visual clues and from the semantic content of the interaction. In
soliloquies (not rare in \textsc{hoc}), the addressee field was left
empty.

\begin{figure*}[!t]
  \subfloat[\textsc{bb}.
  \label{subfig:bb_spk_dur_scene}]{%
    \includegraphics[width=0.3\linewidth]{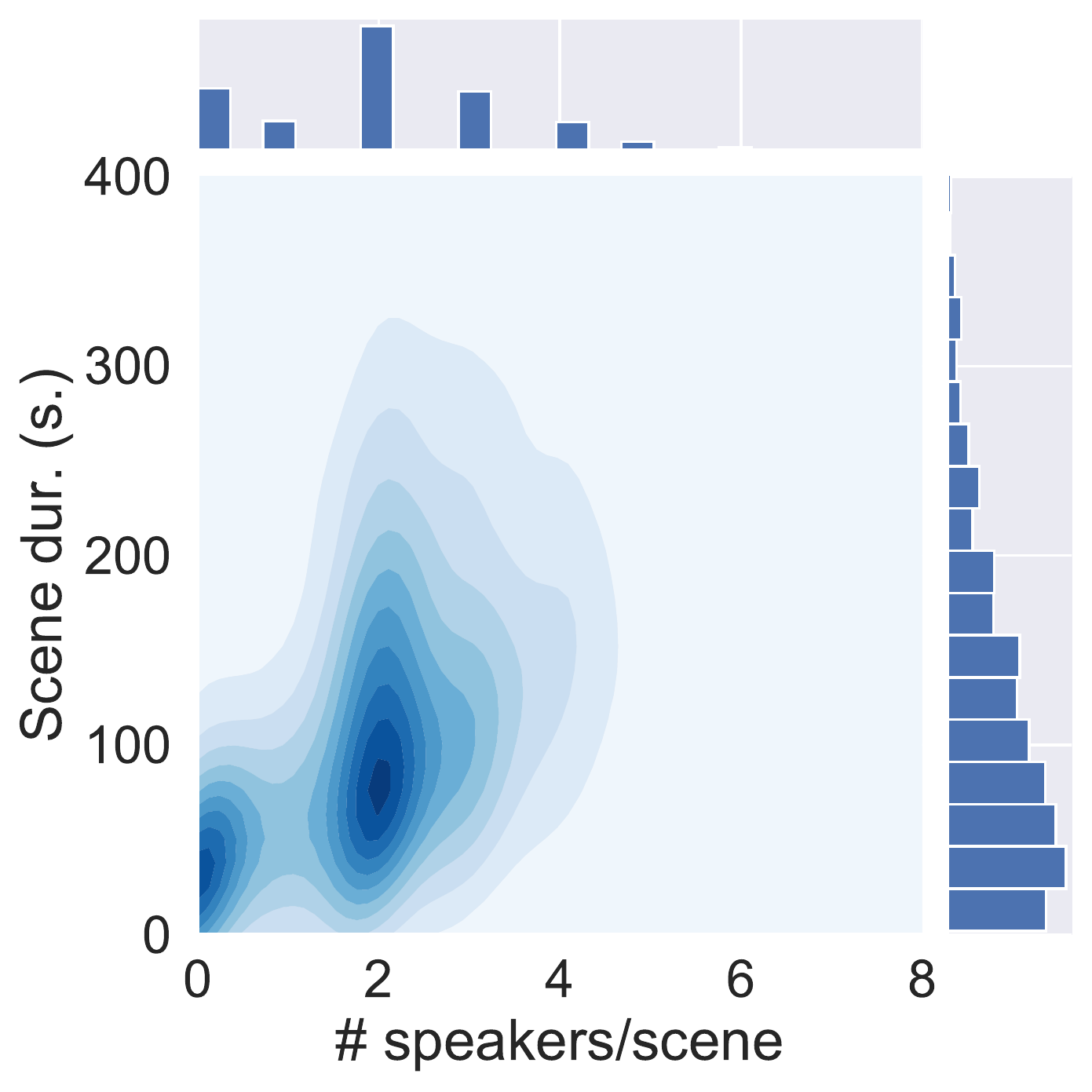}
  }
  \hfill
  \subfloat[\textsc{got}.
  \label{subfig:got_spk_dur_scene}]
  {%
    \includegraphics[width=0.3\linewidth]{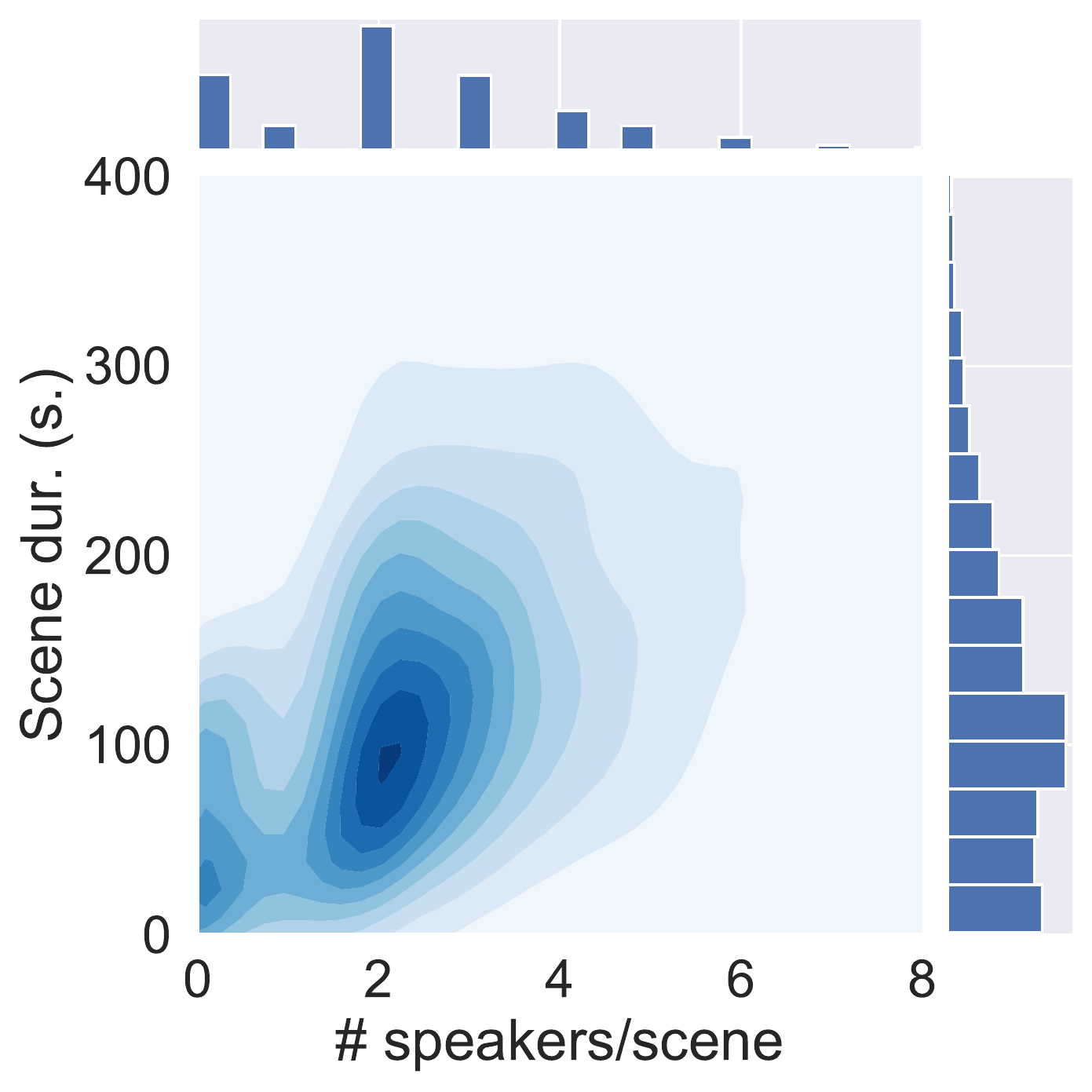}
  }
  \hfill
  \subfloat[\textsc{hoc}.
  \label{subfig:hoc_spk_dur_scene}]
  {%
    \includegraphics[width=0.3\linewidth]{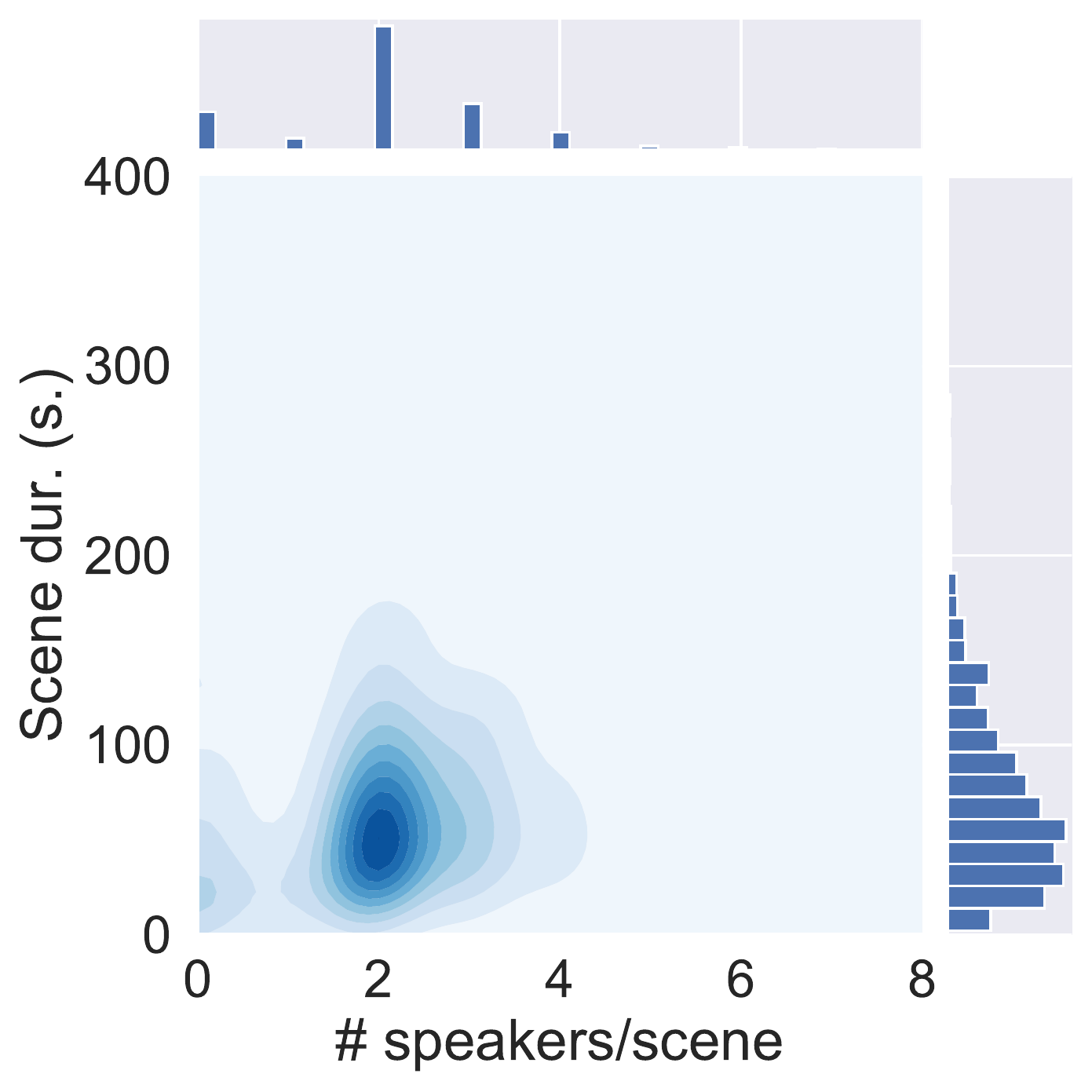}
  }
  \caption{\# speakers/scene \textit{vs.} scene duration.}
  \label{fig:spk_dur_scene}
\end{figure*}

Not frequently addressed alone, the task of determining the
interacting speakers is nonetheless a prerequisite for social
network-based approaches of fiction work analysis, which generally
lack annotated data to intrinsically assess the interactions they
assume \cite{Labatut2019}. Moreover, speaker diarization/recognition
on the one hand, detection of interaction patterns on the other hand,
could probably benefit from one another and be performed jointly. As
an example, Fig.~\ref{fig:networks} shows the conversational networks
based on the annotated episodes for each serial. The vertex sizes
match their degree, while their color corresponds to their betweenness
centrality. This clearly highlights the obvious main characters such
as Walter White (\textsc{bb}) or Francis Underwood (\textsc{hoc}); but
also more secondary characters that have very specific roles
narrative-wise, \textit{e.g.} Jaime Lannister who acts as a bridge
between two groups of characters corresponding to two distinct
narrative arcs. This illustrates the interest of leveraging the social
network of characters when dealing with narrative-related tasks.

\subsection{Shot Boundaries}
\label{subsec:shot_boundaries}

Besides speech oriented annotations, the \textit{Serial Speakers}
dataset contains a few visual annotations. For the first season of each
of the three \textsc{tv} series, we manually annotated shot
boundaries. A video shot, as stated in \cite{Koprinska2001}, is
defined as an ``unbroken sequence of frames taken from one
camera''. Transitions between video shots can be gradual
(fade-in/fade-out), or abrupt ones (cuts). Most of the shot
transitions in our dataset are simple cuts.

The first seasons of \textsc{bb}, \textsc{got}, and \textsc{hoc}
respectively contain 4,416, 9,375 and 8,783 shots, with and average
duration of 4.5, 3.4 and 4.4 seconds. Action scenes in \textsc{got}
are likely to be responsible for shorter shots in average.

Shot boundary detection is nowadays well performed, especially when
consecutive shots are abruptly transitioning from one another. As a
consequence, it is rarely addressed for itself, but as a preliminary
task for more complex ones.

\subsection{Recurring Shots}
\label{subsec:recurring_shots}

\begin{figure}[!h]
 \begin{center}
   \includegraphics[width=0.857\linewidth]{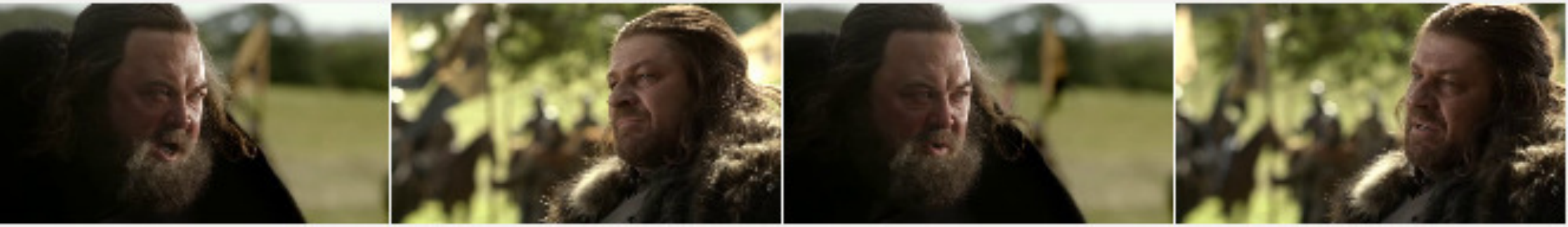} 
   \caption{Example of two alternating recurring shots.}
   \label{fig:recurring_shots}
 \end{center}
\end{figure}

Shots rarely occur only once in edited video streams: in average, a
shot occurs 10 times in \textsc{bb}, 15.2 in \textsc{got} and 17.7 in
\textsc{hoc}. Most of the time, dialogue scenes are responsible for
such shot recurrence. As can be seen on
Fig.~\ref{fig:recurring_shots}, within dialogue scenes, the camera typically
alternates between the interacting characters, resulting in recurring,
possibly alternating, shots.

We manually annotated such recurring shots, based on similar framing,
in the first season of the three \textsc{tv} series. As stated
in~\cite{Yeung1998}, recurring shots usually capture interactions
between characters. Relatively easy to cluster automatically,
recurring shots are especially useful to multimodal approaches
of speaker diarization~\cite{Bost2015}. Besides, recurring shots often
result in complex interaction patterns, denoted \textit{logical story units}
in~\cite{Hanjalic1999}. Such patterns are suitable for supporting local speaker
diarization approaches~\cite{Bost2014}, or for providing extractive
summaries with consistent subsequences~\cite{Bost2019}.


\subsection{Scene Boundaries}
\label{subsec:scene_boundaries}

Scenes are the longest units we annotated in our dataset. As required
by the rule of the three unities classically prescribed for dramas, a
scene in a movie is defined as a homogeneous sequence of actions
occurring at the same place, within a continuous period of time.

Though providing annotators with general guidelines, such a definition
leaves space for interpretation, and some subjective choices still
have to be made to annotate scene boundaries.

First, temporal discontinuity is not always obvious to address:
temporal ellipses often correspond to new scenes, but sometimes,
especially when short, they hardly break the narrative continuity of
the scene.

\begin{figure}[!h]
 \begin{center}
   \includegraphics[width=\linewidth]{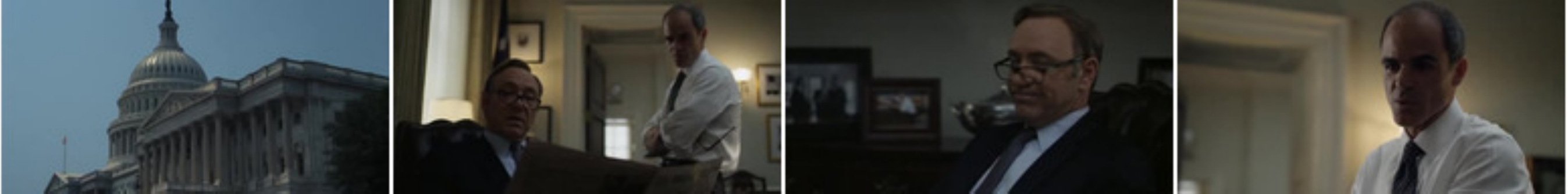} 
   \caption{Long shot opening a scene.}
   \label{fig:long_shot}
 \end{center}
\end{figure}

\begin{table*}[!h]
  \begin{center}
    \input{annot_overview}
    \caption{Annotation overview.}
    \label{tab:overview}
  \end{center}
\end{table*}

Second, as shown on Fig.~\ref{fig:long_shot}, scenes often open with
long shots that show the place of the upcoming scene. Though there is
no, strictly speaking, spatial continuity between the first shot and
the following ones, they obviously belong to the same scene, and
should be annotated as such.

Finally, action homogeneity may also be tricky to assess. For
instance, a phone call within a scene may interrupt an ongoing
dialogue, resulting in a new phone conversation with another
character, and possibly in a new action unit. In such cases, we
generally inserted a new scene to capture the interrupting event, but
other conventions could have been followed. Indeed, the choice of
scene granularity remains highly dependent on the use case the
annotators have in mind for annotating such data: special attention to
speaker interactions would for instance invite to introduce more
frequent scene boundaries.

Overall, \textsc{bb} contains 1,337 scenes, with an average duration
of 127.1 seconds; \textsc{got} 1,813 scenes (avg. duration of 132.6
seconds); \textsc{hoc} 1,048 scenes (avg. duration of 73.2
seconds). Once again, \textsc{hoc} contrasts with the two other
series, with many short scenes.

Fig.~\ref{fig:spk_dur_scene} shows the joint distribution of the
number of speakers by scene and the duration of the scene. For
visualization purposes, the joint distribution is plotted as a
continuous bivariate function, as fitted by applying kernel density
estimate.

As can be seen from the marginal distribution represented horizontally
above each plot, the number of speakers in each scene remains quite
low: 2 in average in \textsc{bb}, 2.1 in \textsc{hoc}, and a bit more
(2.4) in \textsc{got}. Besides, the number of characters in each
scene, except maybe in \textsc{got}, is not clearly correlated with
its duration. Moreover, some short scenes surprisingly do not contain
any speaking character: most of them correspond to the opening and
closing sequences of each episode. Finally, the short scenes of
\textsc{hoc} generally contain two speakers.

Table~\ref{tab:overview} provides an overview of the annotated parts
of the \textit{Serial Speakers} dataset, along with the corresponding
types of annotations. In the table, ``Speech turns'' stand for the
annotation of the speech turns (boundaries, speaker, text); ``Scenes''
for the annotation of the scene boundaries; ``Shots'' for the
annotation of the recurring shots and shot boundaries; and
``Interlocutors'' for the annotation of the interacting speakers\footnote{The annotation files are available online at: \href{https://doi.org/10.6084/m9.figshare.3471839}{doi.org/10.6084/m9.figshare.3471839}}.

\section{Text Recovering Procedure}
\label{sec:textual_content_generation}

Due to copyright restrictions, the published annotation files do not
reproduce the textual content of the speech turns. Instead, the
textual content is encrypted in the public version of the
\textit{Serial Speakers} dataset, and we provide the users with a
simple toolkit to recover the original text from their own subtitle
files\footnote{The toolkit is available online at:
  \href{https://github.com/bostxavier/Serial-Speakers}{github.com/bostxavier/Serial-Speakers}}.

Indeed, the overlap between the textual content of our dataset and the
subtitle files is likely to be large: compared to the annotated text,
subtitles may contain either insertions (formatting tags, sound effect
captions, mentions of speaking characters when not present onscreen),
or some deletions (sentence compression), but very few
substitutions. Every word in the transcript, if not deleted, generally
has the exact same form in the subtitles. As a consequence, the
original word sequence can be recovered from the subtitles. Our text
recovering algorithm first encrypts the tokens found in the subtitle
files provided by the user, before matching the resulting sequence
with the original encrypted token sequence.  The general procedure we
detail below is likely to be of some help to annotators of other movie
datasets with similar copyrighted material.

\subsection{Text Encryption}

For the encryption step, we used truncated hash functions because of the
following desirable properties: deterministic, hash functions
ensure that identical words are encrypted in the same way in the
original text and in the subtitles; they do not reveal information
about the original content, allowing the public version of our dataset
to comply with the copyright restrictions; they are efficient enough
to quickly process the thousands of word types contained in the
subtitles; moreover, once truncated, hash functions result in collisions, able to prevent simple dictionary attacks. Indeed, the main requirement in our case is only to prevent collisions from occurring too close from each other: even if two different words were encrypted in the same way,
they would unlikely be close enough to result in ambiguous
subsequences.

In the public version of our dataset, we compute the first three digits of the SHA-256 hash
function of all of the tokens (including punctuation signs) and
the exact same encryption scheme is applied to the subtitle files, as
provided by the users, resulting in two encrypted token sequences for
every episode of the three \textsc{tv} series.

\subsection{Subtitle Alignment}

We then apply to the two encrypted token sequences the
\textit{Python} \texttt{Difflib} sequence matching
algorithm\footnote{\href{https://docs.python.org/3/library/difflib.html}{docs.python.org/3/library/difflib.html}},
built upon the approach detailed in~\cite{Ratcliff1988}.

Once aligned with the encrypted subtitle sequence, the tokens of the
dataset are decryted by retrieving from the subtitles the original
words.

\begin{figure}[!h]
 \begin{center}
   \includegraphics[width=\linewidth]{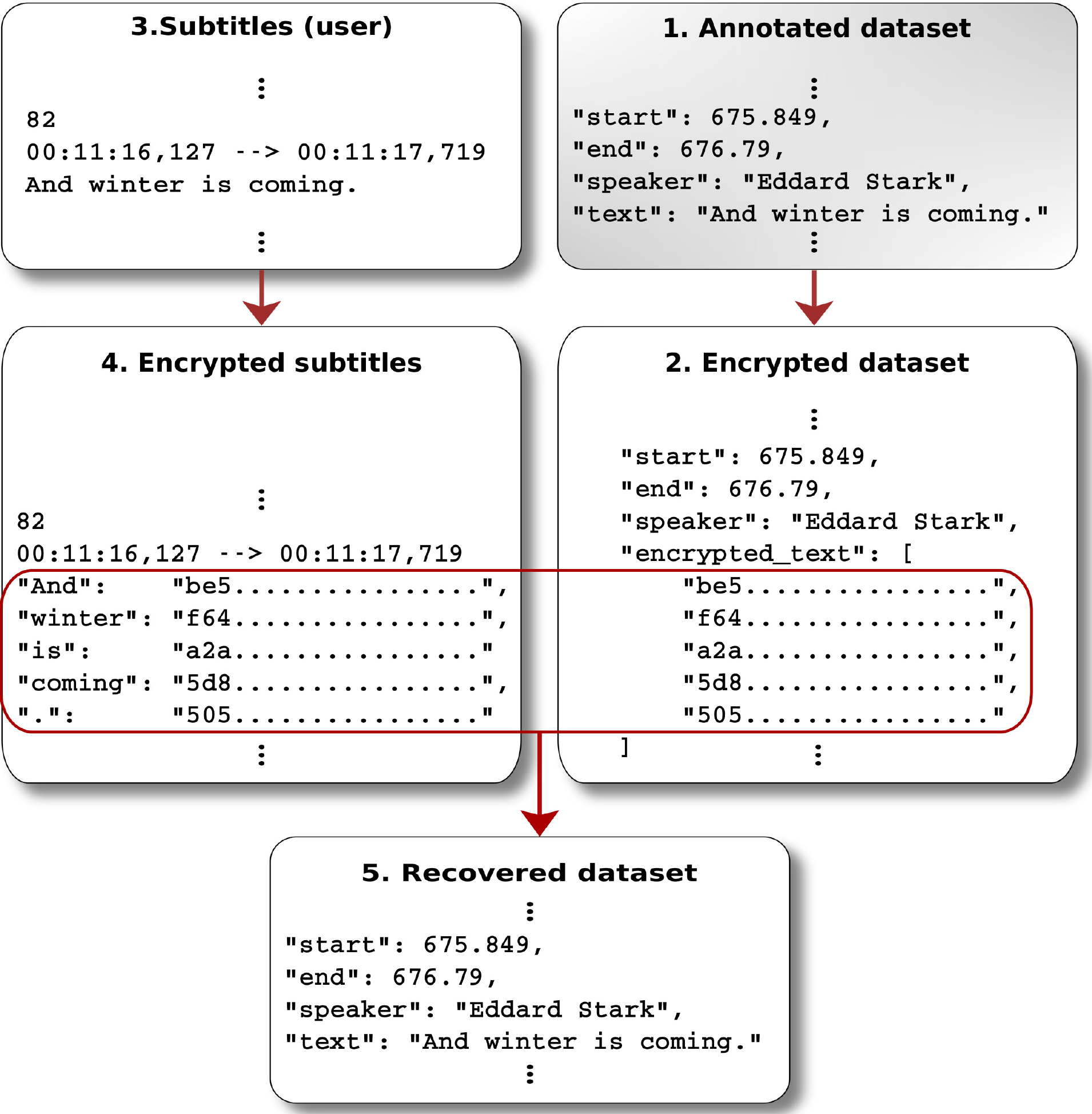} 
   \caption{Text recovering procedure.}
   \label{fig:text_recovering_overview}
 \end{center}
\end{figure}

The whole text recovering procedure is summarized on
Fig.~\ref{fig:text_recovering_overview}. The annotated dataset with
clear text, materialized by the gray box (Box~1) on the figure, is not
publicly available. Instead, in the public annotations, the text is
encrypted (Box~2). In order to recover the text, the user
has to provide h(is/er) own subtitle files (Box~3), which are
encrypted by our tool in the same way as the original dataset text
(Box~4); the resulting encrypted token sequence is matched with
the corresponding token sequence of speech turns (red frame on the
figure), before the text of the speech turns is recovered from the
subtitle words (Box~5).

\subsection{Experiments and Results}

In order to assess the text recovering procedure, we automatically
recovered the textual content from external, publicly available
subtitle files, and compared it to the annotated
text. Table~\ref{tab:text_error_rate} reports in percentage for each
of the three series the average error rates by episode, both computed
at the word level (word error rate, denoted \textsc{wer} in the table)
and at the sentence level (sentence error rate, denoted
\textsc{ser}). In addition, we reported for every episode the average
number of reference tokens (denoted \textit{\# tokens}), and the
average number of insertions, deletions, and substitutions in the
reference word sequence (respectively denoted \textit{Ins},
\textit{Del}, \textit{Sub}). Because of possibly inconsistent
punctuation conventions between the annotated and subtitle text, we
systematically removed the punctuation signs from both sequences
before computing the error rates.

\begin{table}[!h]
  \begin{center}
    \input{text_error_rate}
    \caption{Text recovering: avg. error rates (\%)~/~episode.}
    \label{tab:text_error_rate}
  \end{center}
\end{table}

As can be seen, the average error rates remain remarkably low: the
word error rate amounts to less than 1\% in average. The sentence
error rate also remains quite low: about 1\% for \textsc{got} and
\textsc{hoc}, and a bit higher (4.6\%) for \textsc{bb}. As can be seen
in the right part of the table, deletions are responsible for most of
the errors, especially in \textsc{bb}: as noted in
Subsection~\ref{subsec:speech_turns}, we restored the words missing in
the subtitles when annotating the textual content of the speech
turns. Such missing words turn out to be relatively frequent
in~\textsc{bb}, which can in part explain the higher number of
deletions ($\simeq$ 53 deleted words in average out of $\simeq$
3,700). Moreover, truncating the hash function to the first three
digits does not hurt the performance of the text recovering procedure,
while preventing simple dictionary attacks: the exact same error rates
(not reported in the table) are obtained when keeping the full hash
(64 hexadecimal digits).

In order to allow the user to quickly inspect and edit the differences
between the annotated text and the subtitles, our tool inserts in the
recovered dataset an empty tag \texttt{<>} at the location of deleted
reference tokens. Similarly, we signal every substituted token with an
enclosing tag (\textit{e.g.}  \texttt{<Why>}). As will be seen when
using the toolkit, most of the differences come from different
punctuation/quotation conventions between the annotation and subtitle
files, and rarely impact the vocabulary or the semantics.

The whole recovering process turns out to be fast: 8.3 seconds for
\textsc{got} (73 episodes) on a personal laptop (Intel Xeon-E3-v5
\textsc{cpu}); 6.73 for \textsc{bb} (62 episodes); 4.41 for
\textsc{hoc} (26 episodes). We tried to keep the toolkit as simple as
possible, with a single text recovering \textit{Python} script with few
dependencies.

\section{Conclusion and Perspectives}
\label{sec:conclusion}

In this work, we described \textit{Serial Speakers}, a dataset of 161
annotated episodes from three popular \textsc{tv} serials,
\textit{Breaking Bad} (62 annotated episodes), \textit{Game of
  Thrones} (73), and \textit{House of Cards} (26). \textit{Serial
  Speakers} is suitable for addressing both high level multimedia
retrieval tasks in real world scenarios, and lower level speech
processing tasks in challenging conditions. The boundaries, speaker
and textual content of every speech turn, along with all scene
boundaries, have been manually annotated for the whole set of
episodes; the shot boundaries and recurring shots for the first season of
each of the three series; and the interacting speakers for a subset of 10
episodes. We also detailed the simple text recovering tool we made available to the users, potentially helpful to annotators of other datasets facing similar copyright issues.

As future work, we will first consider including the face
tracks/identities provided for the first season of \textsc{got} in
\cite{Tapaswi2015b}, but these face tracks, automatically generated, would
need manual checking before publication. Furthermore, we plan to
investigate more flexible text encryption schemes: due to the
uniqueness property, hash functions, even truncated, are not tolerant
to spelling/\textsc{ocr} errors in the subtitles. Though the correct
word is generally recovered from the surrounding tokens, it would be
worth investigating encryption functions that would preserve the
similarity between simple variations of the same token.

\section{Acknowledgements}

This work was partially supported by the Research Federation \textit{Agorantic} FR 3621, Avignon University.

\section{Bibliographical References}
\label{main:ref}

\bibliographystyle{lrec}
\bibliography{base}


\end{document}

%% file: speech_features.tex
\begin{tabular}{l|rrr|rrr|rrr}
\toprule
{} & \multicolumn{3}{c}{Speech duration (ratio in \%)} & \multicolumn{3}{c}{\# speech turns} & \multicolumn{3}{c}{\# speakers} \\
\textbf{Show} &                   \textsc{bb} &   \textsc{got} &   \textsc{hoc} &     \textsc{bb} & \textsc{got} & \textsc{hoc} & \textsc{bb} & \textsc{got} & \textsc{hoc} \\
\textbf{Season} &                               &                &                &                 &              &              &             &              &              \\
\midrule
\textbf{1     } &                 02:01:19 (36) &  03:32:55 (40) &  04:50:12 (45) &            4523 &         6973 &        11182 &          59 &          115 &          126 \\
\textbf{2     } &                 03:42:15 (38) &  03:33:53 (41) &  05:07:16 (48) &            8853 &         7259 &        11633 &          86 &          127 &          167 \\
\textbf{3     } &                 03:42:04 (38) &  03:30:01 (39) &        \_ (\_) &            7610 &         7117 &           \_ &          85 &          115 &           \_ \\
\textbf{4     } &                 03:38:08 (37) &  03:11:28 (37) &        \_ (\_) &            7583 &         6694 &           \_ &          70 &          119 &           \_ \\
\textbf{5     } &                 04:40:03 (38) &  02:55:32 (33) &        \_ (\_) &           10372 &         6226 &           \_ &          92 &          121 &           \_ \\
\textbf{6     } &                       \_ (\_) &  02:48:48 (32) &        \_ (\_) &              \_ &         5674 &           \_ &          \_ &          149 &           \_ \\
\textbf{7     } &                       \_ (\_) &  02:13:55 (32) &        \_ (\_) &              \_ &         4526 &           \_ &          \_ &           66 &           \_ \\
\textbf{8     } &                       \_ (\_) &  01:27:17 (21) &        \_ (\_) &              \_ &         3141 &           \_ &          \_ &           50 &           \_ \\
\textbf{Total } &                 17:43:53 (38) &  23:13:52 (35) &  09:57:29 (46) &           38941 &        47610 &        22815 &         288 &          468 &          264 \\
\bottomrule
\end{tabular}

%% file: gen_features.tex
\begin{tabular}{l|rrr}
\toprule
{} & \multicolumn{3}{c}{Duration (\# episodes)} \\
\textbf{Show} &            \textsc{bb} &   \textsc{got} &   \textsc{hoc} \\
\textbf{Season} &                        &                &                \\
\midrule
\textbf{1     } &           05:32:44 (7) &  08:58:28 (10) &  10:48:15 (13) \\
\textbf{2     } &          09:51:08 (13) &  08:41:56 (10) &  10:37:10 (13) \\
\textbf{3     } &          09:49:40 (13) &  08:52:04 (10) &        \_ (\_) \\
\textbf{4     } &          09:46:16 (13) &  08:41:05 (10) &        \_ (\_) \\
\textbf{5     } &          12:15:36 (16) &  08:56:50 (10) &        \_ (\_) \\
\textbf{6     } &                \_ (\_) &  08:55:43 (10) &        \_ (\_) \\
\textbf{7     } &                \_ (\_) &   06:58:54 (7) &        \_ (\_) \\
\textbf{8     } &                \_ (\_) &   06:48:31 (6) &        \_ (\_) \\
\textbf{Total } &          47:15:26 (62) &  66:53:34 (73) &  21:25:26 (26) \\
\bottomrule
\end{tabular}

%% file: annot_overview.tex
\begin{tabular}{l|ccc|ccc|ccc}
\toprule
{} & \multicolumn{3}{c}{Speech turns--Scenes} & \multicolumn{3}{c}{Shots} & \multicolumn{3}{c}{Interlocutors} \\
\textbf{Show} &          \textsc{bb} & \textsc{got} & \textsc{hoc} & \textsc{bb} & \textsc{got} & \textsc{hoc} &   \textsc{bb} & \textsc{got} & \textsc{hoc} \\
\textbf{Season} &                      &              &              &             &              &              &               &              &              \\
\midrule
\textbf{1     } &               \cmark &       \cmark &       \cmark &      \cmark &       \cmark &       \cmark &          4, 6 &      3, 7, 8 &     1, 7, 11 \\
\textbf{2     } &               \cmark &       \cmark &       \cmark &      \xmark &       \xmark &       \xmark &          3, 4 &       \xmark &       \xmark \\
\textbf{3     } &               \cmark &       \cmark &           \_ &      \xmark &       \xmark &           \_ &        \xmark &       \xmark &           \_ \\
\textbf{4     } &               \cmark &       \cmark &           \_ &      \xmark &       \xmark &           \_ &        \xmark &       \xmark &           \_ \\
\textbf{5     } &               \cmark &       \cmark &           \_ &      \xmark &       \xmark &           \_ &        \xmark &       \xmark &           \_ \\
\textbf{6     } &                   \_ &       \cmark &           \_ &          \_ &       \xmark &           \_ &            \_ &       \xmark &           \_ \\
\textbf{7     } &                   \_ &       \cmark &           \_ &          \_ &       \xmark &           \_ &            \_ &       \xmark &           \_ \\
\textbf{8     } &                   \_ &       \cmark &           \_ &          \_ &       \xmark &           \_ &            \_ &       \xmark &           \_ \\
\bottomrule
\end{tabular}

%% file: text_error_rate.tex
\begin{tabular}{l|rr|rrrr}
\toprule
{} &  \textsc{wer} &  \textsc{ser} &  \# tokens &  Ins &   Del &  Sub \\
\textbf{Show        } &               &               &            &      &       &      \\
\midrule
\textbf{\textsc{bb} } &           1.6 &           4.6 &     3699.8 &  0.2 &  53.1 &  4.1 \\
\textbf{\textsc{got}} &           0.4 &           1.2 &     4353.2 &  0.1 &  13.8 &  1.9 \\
\textbf{\textsc{hoc}} &           0.2 &           0.7 &     5918.0 &  0.1 &   7.4 &  2.3 \\
\bottomrule
\end{tabular}

%% file: serial_speakers.bbl
\begin{thebibliography}{}

\bibitem[\protect\citename{Barthélemy \bgroup et al.\egroup
  }2005]{Barthelemy2005}
Barthélemy, M., Barrat, A., Pastor-Satorras, R., and Vespignani, A.
\newblock (2005).
\newblock Characterization and modeling of weighted networks.
\newblock {\em Physica A}, 346(1-2):34--43.

\bibitem[\protect\citename{B{\"a}uml \bgroup et al.\egroup }2013]{Bauml2013}
B{\"a}uml, M., Tapaswi, M., and Stiefelhagen, R.
\newblock (2013).
\newblock Semi-supervised learning with constraints for person identification
  in multimedia data.
\newblock In {\em IEEE Conference on Computer Vision and Pattern Recognition},
  pages 3602--3609.

\bibitem[\protect\citename{B{\"a}uml \bgroup et al.\egroup }2014]{Bauml2014}
B{\"a}uml, M., Tapaswi, M., and Stiefelhagen, R.
\newblock (2014).
\newblock A time pooled track kernel for person identification.
\newblock In {\em 11th IEEE International Conference on Advanced Video and
  Signal Based Surveillance}, pages 7--12.

\bibitem[\protect\citename{Bost and Linares}2014]{Bost2014}
Bost, X. and Linares, G.
\newblock (2014).
\newblock Constrained speaker diarization of tv series based on visual
  patterns.
\newblock In {\em IEEE Spoken Language Technology Workshop}, pages 390--395.

\bibitem[\protect\citename{Bost \bgroup et al.\egroup }2015]{Bost2015}
Bost, X., Linar{\`e}s, G., and Gueye, S.
\newblock (2015).
\newblock Audiovisual speaker diarization of tv series.
\newblock In {\em 2015 IEEE International Conference on Acoustics, Speech and
  Signal Processing}, pages 4799--4803. IEEE.

\bibitem[\protect\citename{Bost \bgroup et al.\egroup }2019]{Bost2019}
Bost, X., Gueye, S., Labatut, V., Larson, M., Linar{\`e}s, G., Malinas, D., and
  Roth, R.
\newblock (2019).
\newblock Remembering winter was coming.
\newblock {\em Multimedia Tools and Applications}, 78(24):35373--35399, Dec.

\bibitem[\protect\citename{Bost}2016]{Bost2016}
Bost, X.
\newblock (2016).
\newblock {\em A storytelling machine? Automatic video summarization: the case
  of TV series}.
\newblock {Ph.D.} thesis.

\bibitem[\protect\citename{Bredin and Gelly}2016]{Bredin2016}
Bredin, H. and Gelly, G.
\newblock (2016).
\newblock Improving speaker diarization of tv series using talking-face
  detection and clustering.
\newblock In {\em 24th ACM international conference on Multimedia}, pages
  157--161.

\bibitem[\protect\citename{Bredin}2012]{Bredin2012}
Bredin, H.
\newblock (2012).
\newblock Segmentation of tv shows into scenes using speaker diarization and
  speech recognition.
\newblock In {\em IEEE International Conference on Acoustics, Speech and Signal
  Processing}, pages 2377--2380.

\bibitem[\protect\citename{Clauset \bgroup et al.\egroup }2009]{Clauset2009}
Clauset, A., Shalizi, C.~R., and Newman, M. E.~J.
\newblock (2009).
\newblock Power-law distributions in empirical data.
\newblock {\em SIAM Review}, 51(4):661--703.

\bibitem[\protect\citename{Cl{\'e}ment \bgroup et al.\egroup
  }2011]{Clement2011}
Cl{\'e}ment, P., Bazillon, T., and Fredouille, C.
\newblock (2011).
\newblock Speaker diarization of heterogeneous web video files: A preliminary
  study.
\newblock In {\em IEEE International Conference on Acoustics, Speech and Signal
  Processing}, pages 4432--4435.

\bibitem[\protect\citename{Ercolessi \bgroup et al.\egroup
  }2011]{Ercolessi2011}
Ercolessi, P., Bredin, H., S{\'e}nac, C., and Joly, P.
\newblock (2011).
\newblock Segmenting tv series into scenes using speaker diarization.
\newblock In {\em Workshop on Image Analysis for Multimedia Interactive
  Services}, pages 13--15.

\bibitem[\protect\citename{Ercolessi \bgroup et al.\egroup
  }2012a]{Ercolessi2012b}
Ercolessi, P., Bredin, H., and S{\'e}nac, C.
\newblock (2012a).
\newblock Stoviz: story visualization of tv series.
\newblock In {\em 20th ACM international conference on Multimedia}, pages
  1329--1330.

\bibitem[\protect\citename{Ercolessi \bgroup et al.\egroup
  }2012b]{Ercolessi2012a}
Ercolessi, P., S{\'e}nac, C., and Bredin, H.
\newblock (2012b).
\newblock Toward plot de-interlacing in tv series using scenes clustering.
\newblock In {\em 10th International Workshop on Content-Based Multimedia
  Indexing}, pages 1--6.

\bibitem[\protect\citename{Everingham \bgroup et al.\egroup
  }2006]{Everingham2006}
Everingham, M., Sivic, J., and Zisserman, A.
\newblock (2006).
\newblock Hello! my name is... buffy''--automatic naming of characters in tv
  video.
\newblock In {\em BMVC}, volume~2, page~6.

\bibitem[\protect\citename{Friedland \bgroup et al.\egroup
  }2009]{Friedland2009}
Friedland, G., Gottlieb, L., and Janin, A.
\newblock (2009).
\newblock Using artistic markers and speaker identification for narrative-theme
  navigation of seinfeld episodes.
\newblock In {\em 11th IEEE International Symposium on Multimedia}, pages
  511--516.

\bibitem[\protect\citename{Ghaleb \bgroup et al.\egroup }2015]{Ghaleb2015}
Ghaleb, E., Tapaswi, M., Al-Halah, Z., Ekenel, H.~K., and Stiefelhagen, R.
\newblock (2015).
\newblock {Accio: A Data Set for Face Track Retrieval in Movies Across Age}.
\newblock In {\em ACM International Conference on Multimedia Retrieval}.

\bibitem[\protect\citename{Hanjalic \bgroup et al.\egroup }1999]{Hanjalic1999}
Hanjalic, A., Lagendijk, R.~L., and Biemond, J.
\newblock (1999).
\newblock Automated high-level movie segmentation for advanced video-retrieval
  systems.
\newblock {\em IEEE Transactions on Circuits and Systems for Video Technology},
  9(4):580--588.

\bibitem[\protect\citename{Koprinska and Carrato}2001]{Koprinska2001}
Koprinska, I. and Carrato, S.
\newblock (2001).
\newblock Temporal video segmentation: A survey.
\newblock {\em Signal processing: Image communication}, 16(5):477--500.

\bibitem[\protect\citename{Labatut and Bost}2019]{Labatut2019}
Labatut, V. and Bost, X.
\newblock (2019).
\newblock Extraction and analysis of fictional character networks: A survey.
\newblock {\em ACM Computing Surveys}, 52(5):89.

\bibitem[\protect\citename{Li and Chen}2003]{Li2003a}
Li, C. and Chen, G.
\newblock (2003).
\newblock Network connection strengths: Another power-law?
\newblock {\em arXiv}, cond-mat.dis-nn:0311333.

\bibitem[\protect\citename{McAuliffe \bgroup et al.\egroup
  }2017]{Mcauliffe2017}
McAuliffe, M., Socolof, M., Mihuc, S., Wagner, M., and Sonderegger, M.
\newblock (2017).
\newblock Montreal forced aligner: Trainable text-speech alignment using kaldi.
\newblock In {\em Interspeech}, pages 498--502.

\bibitem[\protect\citename{McCarthy and Jarvis}2010]{McCarthy2010}
McCarthy, P.~M. and Jarvis, S.
\newblock (2010).
\newblock Mtld, vocd-d, and hd-d: A validation study of sophisticated
  approaches to lexical diversity assessment.
\newblock {\em Behavior research methods}, 42(2):381--392.

\bibitem[\protect\citename{Ratcliff and Metzener}1988]{Ratcliff1988}
Ratcliff, J.~W. and Metzener, D.~E.
\newblock (1988).
\newblock Pattern-matching-the gestalt approach.
\newblock {\em Dr Dobbs Journal}, 13(7):46.

\bibitem[\protect\citename{Roy \bgroup et al.\egroup }2014]{Roy2014}
Roy, A., Guinaudeau, C., Bredin, H., and Barras, C.
\newblock (2014).
\newblock Tvd: a reproducible and multiply aligned tv series dataset.
\newblock In {\em 9th International Conference on Language Resources and
  Evaluation}, page 418–425.

\bibitem[\protect\citename{Tapaswi \bgroup et al.\egroup }2012]{Tapaswi2012}
Tapaswi, M., Bäuml, M., and Stiefelhagen, R.
\newblock (2012).
\newblock {``Knock! Knock! Who is it?'' Probabilistic Person Identification in
  TV series}.
\newblock In {\em IEEE Conference on Computer Vision and Pattern Recognition}.

\bibitem[\protect\citename{Tapaswi \bgroup et al.\egroup }2014a]{Tapaswi2014b}
Tapaswi, M., Bäuml, M., and Stiefelhagen, R.
\newblock (2014a).
\newblock {Story-based Video Retrieval in TV series using Plot Synopses}.
\newblock In {\em ACM International Conference on Multimedia Retrieval}.

\bibitem[\protect\citename{Tapaswi \bgroup et al.\egroup }2014b]{Tapaswi2014a}
Tapaswi, M., Bäuml, M., and Stiefelhagen, R.
\newblock (2014b).
\newblock {StoryGraphs: Visualizing Character Interactions as a Timeline}.
\newblock In {\em IEEE Conference on Computer Vision and Pattern Recognition}.

\bibitem[\protect\citename{Tapaswi \bgroup et al.\egroup }2015a]{Tapaswi2015b}
Tapaswi, M., Bäuml, M., and Stiefelhagen, R.
\newblock (2015a).
\newblock {Book2Movie: Aligning Video scenes with Book chapters}.
\newblock In {\em IEEE Conference on Computer Vision and Pattern Recognition}.

\bibitem[\protect\citename{Tapaswi \bgroup et al.\egroup }2015b]{Tapaswi2015a}
Tapaswi, M., Bäuml, M., and Stiefelhagen, R.
\newblock (2015b).
\newblock {Improved Weak Labels using Contextual Cues for Person Identification
  in Videos}.
\newblock In {\em IEEE International Conference on Automatic Face and Gesture
  Recognition}.

\bibitem[\protect\citename{Tran \bgroup et al.\egroup }2011]{Tran2011}
Tran, V.-A., Le, V., Barras, C., and Lamel, L.
\newblock (2011).
\newblock Comparing multi-stage approaches for cross-show speaker diarization.

\bibitem[\protect\citename{Yeung \bgroup et al.\egroup }1998]{Yeung1998}
Yeung, M., Yeo, B.-L., and Liu, B.
\newblock (1998).
\newblock Segmentation of video by clustering and graph analysis.
\newblock {\em Computer vision and image understanding}, 71(1):94--109.

\end{thebibliography}
